\documentclass[preprint,amsmath,amssymb,superscriptaddress, floatfix,nofootinbib,notitlepage]
{revtex4}
\usepackage{epsfig,subfigure}
\usepackage{epstopdf}
\usepackage[utf8]{inputenc}
\usepackage{graphicx}
\usepackage{amssymb}
\usepackage{amsxtra}
\usepackage{amsmath,amstext}
\usepackage{booktabs,multirow,tabularx}
\usepackage{slashed}
\usepackage{float}
\usepackage{placeins}
\usepackage{color}
\usepackage{hyperref}
\usepackage{soul}
\usepackage{braket}
\usepackage{physics}
\usepackage[absolute]{textpos}
\usepackage[T1]{fontenc}
\usepackage[super]{nth}
\usepackage{enumitem}
\usepackage{listings,lipsum}
\usepackage{xcolor}
\usepackage{xspace}
\usepackage{soul}
\usepackage{tabularx}
\definecolor{codegreen}{rgb}{0,0.6,0}
\definecolor{codegray}{rgb}{0.5,0.5,0.5}
\definecolor{codepurple}{rgb}{0.58,0,0.82}
\definecolor{backcolour}{rgb}{0.95,0.95,0.92}
\lstdefinestyle{mystyle}{
backgroundcolor=\color{backcolour},      commentstyle=\color{codegreen},    keywordstyle=\color{magenta},  numberstyle=\tiny\color{codegray},
 stringstyle=\color{codepurple}, basicstyle=\ttfamily\footnotesize,
    breakatwhitespace=false,     
    breaklines=true,             
    captionpos=b,                
    keepspaces=true,             
    numbers=left,                
    numbersep=5pt,               
    showspaces=false,            
    showstringspaces=false,
    showtabs=false,              
    tabsize=2
}
\newcommand{\mg}{\textsc{MG5aMC}\xspace}
\newcommand{\fr}{\textsc{FeynRules}\xspace}
\newcommand{\HELAS}{\textsc{Helas}\xspace}
\newcommand{\ALOHA}{\textsc{Aloha}\xspace}
\newcommand{\UFO}{\textsc{Ufo}\xspace}
\newcommand{\rts}{\sqrt{s}}

\begin{document}
 \begin{flushright}
 KEK-TH-2616
 \quad
 IRMP-CP3-24-12
 \end{flushright}
 
\title{Automatic generation of helicity amplitudes\\ in Feynman-Diagram gauge}
\author{Kaoru Hagiwara}
\email{kaoru.hagiwara@kek.jp}
\address{KEK Theory Center and Sokendai, Tsukuba, Ibaraki 305-0801, Japan}
\author{Junichi Kanzaki}
\email{kanzaki@post.kek.jp}
\address{KEK Theory Center and Sokendai, Tsukuba, Ibaraki 305-0801, Japan}
\author{Olivier~Mattelaer}
\email{olivier.mattelaer@uclouvain.be}
\address{CP3, Université Catholique de Louvain, Chemin du Cyclotron, B-1348 Louvain la Neuve, Belgium}
\author{Kentarou Mawatari }
\email{mawatari@iwate-u.ac.jp}
\address{Faculty of Education,  Iwate University, Morioka, Iwate 020-8550, Japan}
\address{Graduate School of Arts and Sciences, Graduate School of Science and Engineering, Iwate University, Morioka, Iwate 020-8550, Japan}
\author{Ya-Juan Zheng}
\email{yjzheng@iwate-u.ac.jp}
\address{Faculty of Education, Iwate University, Morioka, Iwate 020-8550, Japan}
\begin{abstract}
We develop a method to calculate
helicity amplitudes of an arbitrary tree-level process in
Feynman-Diagram (FD) gauge for an arbitrary gauge model with {MadGraph5\_aMC@NLO}.
We start from the 't Hooft--Feynman gauge Lagrangian in
\fr~and generate scattering amplitudes by identifying 
the Goldstone boson as the $\nth{5}$ component of each weak boson.
All the vertices of the 5-component weak bosons are then
created automatically by assembling the relevant weak boson 
and Goldstone boson vertices in the Feynman gauge.
The 5-component weak boson vertices are then connected by the
$5 \times 5$ matrix propagator in the FD gauge. 
As a demonstration of the method we calculate the cross
section for the process
$ \mu^- \mu^+ \to \nu_\mu \bar{\nu}_\mu t \bar{t} H $
with complex top Yukawa coupling, which can be obtained
by adding a gauge invariant dimension-6 operator to
the Standard Model (SM) Lagrangian.
The FD gauge and the unitary (U) gauge amplitudes give
exactly the same cross section, and subtle gauge theory
cancellation among diagrams in the U gauge at high energies
is absent in the FD gauge, as has been observed for
various SM processes.
In addition, we find that the total cross sections at high energies are dominated by a single, or a set of non-vanishing Feynman amplitudes with the higher dimensional vertices in the FD gauge.
\end{abstract}
\maketitle
\section{Introduction}
Recently a new
gauge boson propagator has been proposed
for massless gauge bosons~~\cite{Hagiwara:2020tbx} and for massive gauge
bosons~~\cite{Chen:2022gxv, Chen:2022xlg}.
They can be obtained from the light cone (LC) gauge~\cite{Chen:2022xlg},
where the gauge vector is chosen along the opposite of
the gauge boson three momentum direction:
\begin{eqnarray}
n(q)_{\rm FD}^\mu = ( {\rm sgn}(q^0), -\vec{q}/|\vec{q}| )
.
\label{eq:FD}
\end{eqnarray}
This particular choice of the LC gauge vector has
been named Feynman-Diagram (FD) gauge~~\cite{Chen:2022gxv, Chen:2022xlg}, because
of the common feature that subtle cancellation among
interfering Feynman diagrams is absent~~\cite{Hagiwara:2020tbx,Chen:2022gxv}, that the
observable cross sections are dominated by a single
Feynman diagram in the singular kinematical configuration
where the Parton Shower description holds~~\cite{Hagiwara:2020tbx,Chen:2022gxv,Hagiwara:2009wt,Wulzer:2013mza,Chen:2016wkt},
and that some interference patterns away from the singular
region seem to allow physical interpretation, such
as the angular ordering of QCD radiations~~\cite{Chen:2022xlg,Nagy:2007ty,Nagy:2014mqa}.

Because the subtle cancellation among interfering
Feynman diagrams has been a severe obstruct in
numerical evaluation of the amplitudes~~\cite{Hagiwara:1990gk,Ruiz:2021tdt},
it is desirable that all numerical codes for
the scattering amplitudes be available in the
FD gauge.
In QED and QCD, this is readily available since the
only necessary change is to replace the photon and
gluon propagators, which are given in the Feynman
gauge in \HELAS~~\cite{Hagiwara:1990dw,Murayama:1992gi} adopted by 
the series of {\textsc MadGraph}~~\cite{Stelzer:1994ta,Alwall:2007st,Alwall:2011uj,Alwall:2014hca},
one of the commonly-used matrix-element event generators.
In the electroweak (EW) sector, we need to introduce
new Feynman rules which treat the Goldstone bosons
as the \nth{5} component of the massive weak bosons.

In the paper~~\cite{Chen:2022gxv} , the 5$\times$5 FD gauge propagator of
massive weak bosons~~\cite{Chen:2022gxv,Wulzer:2013mza} has been derived from the
unitary gauge propagator by making use of the
BRST identities~~\cite{Becchi:1975nq,Tyutin:1975qk},
which relates the off-shell
amplitudes of the scalar component of the
weak boson $\partial_\mu V^\mu$ and that of the
corresponding Goldstone boson $\pi_V$.
It has been found in ref.~~\cite{Chen:2022gxv} that we need to introduce three
new vertices, $ZZZZ$, $WWZH$ and $WWAH$,
which do not appear in the Feynman rules of the
Standard Model (SM) in the unitary gauge.
In ref.~~\cite{Chen:2022xlg}, the FD gauge propagator has been
obtained directly as the Green's function of
the equations of motion (EOM) of the LC gauge quantized EW theory.
In this representation, it is clear that the
Goldstone bosons are the \nth{5} component of
the weak boson, since the EOM mix all the five
components.

In ref.~~\cite{Chen:2022gxv}, the FD gauge Feynman rules of the SM are
obtained by supplying the above three 4-point vertices
to the standard unitary gauge vertices, and then all
the vertex functions among \nth{5} components of the weak bosons
have been coded manually.
In ref.~~\cite{Chen:2022xlg}, the Feynman rules are obtained directly
from the Lagrangian, since the Goldstone bosons are
the \nth{5} component of the massive weak bosons.
Therefore, we already have the complete Feynman rules
and the vertex function programs to calculate the
tree-level scattering amplitudes in {MadGraph5\_aMC@NLO} (\mg)~~\cite{Alwall:2014hca} for
an arbitrary process in the SM.

In this paper, we would like to extend the coverage
to an arbitrary gauge model, where all the massive
gauge bosons are obtained by the spontaneous breaking
of gauge symmetries in the scalar (Higgs) potential.
We start from the \fr~~\cite{Alloul:2013bka} model in the 't~Hooft--Feynman
gauge, and the Feynman rules and the corresponding
vertex functions for numerical calculations of scattering
amplitudes are automatically generated~~\cite{Degrande:2011ua,Darme:2023jdn,deAquino:2011ub} in \mg\ for 
an arbitrary gauge model. 
The key observation of ours is that all the gauge boson
and the Goldstone boson vertices are gauge invariant,
since the gauge dependence appears only in the propagators,
i.e. in the weak boson and the Goldstone boson propagators
in the covariant $R_\xi$ gauge~\cite{Fujikawa:1972fe}.
Therefore, all the vertex functions (the~\HELAS~codes)
generated in the Feynman gauge are valid also in
the FD gauge.
Our task is hence to combine them into the vertex
function among the 5-component weak bosons, automatically.
Once this is done, the $5\times5$ weak boson propagators connecting
two 5-component vertices should be the FD gauge propagators.

The paper is organized as follows:
In section~\ref{sec:2}, we explain how the weak boson and the
Goldstone boson vertex functions (the \HELAS~codes) should
be combined to make a new vertex function
among the 5-component weak boson, automatically.
In section~\ref{sec:3}, we apply the method to a simplest extension
of the SM, the SM effective field theory (SMEFT)~\cite{Leung:1984ni,Buchmuller:1985jz,Grzadkowski:2010es} with a
single dimension-6 operator.
We select an operator which modifies the top quark Yukawa
coupling to the Higgs boson, making it CP violating~\cite{Zhang:1994fb,Whisnant:1994fh,Chen:2022yiu,Liu:2023yrb,Barger:2023wbg,Cassidy:2023lwd}.
We show that the absence of subtle cancellation among
interfering Feynman diagrams persists in the presence
of non-Standard interactions.
Section~\ref{sec:4} summarises our findings.
%
%%%%%%%%%%%%%%%%%%%%%%%%%%%%%%%%%%%%%%%%%%%%%%%%%%%%%
\section{From Feynman gauge to Feynman-Diagram gauge}
\label{sec:2}
In the original FD gauge papers~\cite{Hagiwara:2020tbx,Chen:2022gxv,Chen:2022xlg}, the amplitudes in the FD gauge have been calculated by using an independent \HELAS~ library, which was coded manually and specifically prepared for all the SM vertices.%
However, in order to aid our pursuits for physics beyond the SM, it is necessary to automatically generate the codes to calculate FD gauge amplitudes for any processes in arbitrary gauge models.
In this section, we show how this can be achieved in the framework of \mg.

Since \mg\ supports calculations in the Feynman gauge, it includes all the elements needed to incorporate calculations in the FD gauge.
We, therefore, introduce a new command in \mg\ in order to let the user change the gauge;
\begin{small}
\begin{lstlisting}[
language=C,
backgroundcolor=\color{backcolour}]
set gauge ***
\end{lstlisting}
\end{small}
where {\tt ***} can be {\tt Feynman}, {\tt unitary}, {\tt axial} or {\tt FD} for the FD gauge.
Accordingly, we can now study any processes of interest in the FD gauge with only a few \mg\ commands.%
\footnote{Note that the present FD gauge option in \mg\ is restricted to LO/tree-level processes, both in standalone mode (evaluation of amplitude only) and in madevent mode (computation of cross-section and generation of unweighted events).} 
For instance, for the $ZZ\to ZZ$ process:
%\begin{verbatim}
%    import model sm
%    set gauge FD
  
%    generate z z > z z
%    output
%    launch
%\end{verbatim}
\begin{small}
\begin{lstlisting}[
language=C,
backgroundcolor=\color{backcolour}]
MG5_aMC>import model sm 
MG5_aMC>set gauge FD 
MG5_aMC>generate z z > z z
MG5_aMC>output
MG5_aMC>launch
\end{lstlisting}    
\end{small}
Without "{\tt set gauge ***}", \mg\ employs the unitary gauge as a default gauge choice for massive gauge bosons. 
For the above process, there are three Higgs-exchanged diagrams in the unitary gauge.
On the other hand, one additional four-point contact diagram exists in the FD gauge. 
We refer to Sec.~3.1 in ref.~\cite{Chen:2022gxv} for more details.

The command "{\tt set gauge FD}" internally modifies both the representation of the model within \mg\ and the way the \ALOHA\ \cite{deAquino:2011ub} code generates the \HELAS\ subroutines. 
We describe each modification step-by-step below.

\begin{description}[labelwidth=\linewidth]
\item[(i) Extend vector boson wave functions from 4 components to 5 ones]

Since the vector boson wave function (or polarization vector) for massive gauge bosons in the FD gauge includes the Goldstone boson component, 
the usual 4-component wave function $\epsilon^\mu(p,\lambda)$ must be extended to the 5-component one $\epsilon^M(p,\lambda)$,
where $p$ and $\lambda\,(=\!\!\pm1,0)$ are the momentum and the helicity of the vector boson, respectively, and the index $M$ runs from 0 to 4, $M=\{\mu,4\}$.
  
The \HELAS\ vector boson wave function is a one-dimensional array of double complex numbers, with a four-momentum $p^\mu$ defined in the first two slots of the wave function variable, \texttt{VC}:
%\begin{align}
%\mathtt{(VC(1),VC(2)) = NSV\, ( p(0) + \mathit{i}\, p(3), p(1) + \mathit{i}\, p(2))}.
%\end{align}
%
\begin{small}
\begin{lstlisting}[
language=Python,
backgroundcolor=\color{backcolour}]
(VC(1), VC(2)) = NSV (p(0) + i p(3), p(1) + i p(2))
\end{lstlisting}
\end{small}
An integer, \texttt{NSV}, is defined as \texttt{NSV=+1} if the vector boson is in the final state and \texttt{NSV=-1} if it is in the initial state.

Normally, the following four complex numbers correspond to the vector boson wave function $\epsilon^\mu(p,\lambda)$: 
\begin{small}
\begin{lstlisting}[
language=Fortran,
backgroundcolor=\color{backcolour}]
(VC(3), VC(4), VC(5), VC(6))
\end{lstlisting}
\end{small}  
This must be changed to the 5-component wave function $\epsilon^M(p,\lambda)$
to include the Goldstone boson component as
\begin{small}
\begin{lstlisting}[
language=Python,
gobble=-2,
backgroundcolor=\color{backcolour}]
(VC(3), VC(4), VC(5), VC(6), VC(7))
\end{lstlisting}
\end{small}
Appendix~A.2 of ref.~\cite{Chen:2022gxv} includes the complete code for defining the vector boson wave function in the FD gauge.

\item[(ii) Identify Goldstone bosons as their associated gauge bosons]
\UFO\ models~\cite{Degrande:2011ua,Darme:2023jdn}  compatible with the Feynman gauge contain Goldstone bosons and their associated interactions. 
With the command "{\tt set gauge FD}", \mg\ automatically, based on the mass, identify which Goldstone boson is associated with which vector boson, e.g. ($\pi^\pm,\pi^0$) as  ($W^\pm,Z$) in the SM.
%\footnote{In this section we follow the \mg\ notation ($G^\pm,G^0$) for Goldstone bosons, otherwise we use ($\pi^\pm,\pi^0$).}

In each interaction vertex, all the Goldstone bosons are replaced by their associated vector bosons, 
while the color and Lorentz structures of the interaction are kept untouched.%
\footnote{This can lead to a warning due to the fact that a Lorentz structure for a scalar is attached to a spin-one particle.}
If an interaction for the same particle content exists, the two interactions are merged into a single one.

For example, the interaction $W^+\pi^-H$ present below in the \UFO\ format:%
\footnote{We do not exactly follow the \UFO\ format to make it more readable/understandable for the reader.}

\begin{small}
\begin{lstlisting}[language=Python, 
backgroundcolor=\color{backcolour}]
particles: [24,-251,25], #  W+ Pi- H
color: [c0 = 1 ],
lorentz: [VSS1 = P(1,2) - P(1,3)],
couplings: {(c0, VSS1): GC_37=-ee/(2.*sw)},
orders: {QED: 1},
\end{lstlisting}
\end{small}
will be merged with the interaction $W^+W^-H$: 
\begin{small}
\begin{lstlisting}[language=Python,
backgroundcolor=\color{backcolour}]
particles: [-24,24,25], # W- W+ H 
color: [c0 = 1 ],
lorentz: [VVS1 = Metric(1,2)],
couplings: {(c0, VVS1): GC_72=(ee**2*complexi*vev)/(2.*sw**2)},
orders: {QED: 1},
\end{lstlisting}
\end{small}
One can note that the position of the $W^+$ particle is not the same between the two interactions, and therefore, one needs to adapt the definition of the Lorentz (and, in principle, color) structure to the new ordering before actually doing the merge. 
The permutation of the indices is done automatically, and the new Lorentz and color structures are defined when/if needed (which, in this case, created the new SVS1 Lorentz structure). 
This procedure gives the following merge interaction:%
\footnote{Additional interactions like $\pi^+W^-H$ will also be merged into the same interaction, but those are not shown here for clarity.
We refer to Table~4 in ref.~\cite{Chen:2022gxv} for more details.
}

\begin{small}
\begin{lstlisting}[language=Python,
backgroundcolor=\color{backcolour}]
particles: [-24,24,25], # W- W+ H
color: [c0 = 1 ],
lorentz: [VVS1 = Metric(1,2), 
          SVS1 = P(2,1) - P(2,3)],
couplings: {(c0, VVS1): GC_72=(ee**2*complexi*vev)/(2.*sw**2),
            (c0, SVS1): GC_37=-ee/(2.*sw)},
orders: {QED: 1},
\end{lstlisting}
\end{small}

One technical difficulty arises for interactions where, in the Feynman gauge, both the vector boson and its Goldstone boson counterpart are present. To make the issue clearer, we present an example for the case of the four-point interaction among two $Z$ bosons and two of their Goldstone bosons in the SM: %In the \UFO\ format
%\begin{verbatim}
\begin{small}
\begin{lstlisting}[language=Python,
backgroundcolor=\color{backcolour}]
particles: [23,23,250,250], # Z Z Pi0 Pi0
color: [c0 = 1 ],
lorentz: [VVSS1 = Metric(1,2)],
couplings: {(c0, VVSS1): GC_65=ee**2*i + (cw**2*ee**2*i)/(2*sw**2) 
                                 + (ee**2*i*sw**2)/(2*cw**2)},
orders: {QED: 2},
\end{lstlisting}
\end{small}
%\end{verbatim}
%
As described above, the interaction will be mapped into a new interaction with four $Z$ bosons, which is not present neither in the Feynman gauge nor in the unitary gauge. The issue is that any pair of $Z$ bosons need to be associated with a new Lorentz structure where they correspond to a Goldstone boson, leading to the final interaction containing six Lorentz structures.
Five of those Lorentz structures are new and are generated automatically, given that they are identical to the original one up to the permutation of the indices.
The new interaction is then given by:%
\footnote{In the SM, a \nth{7} Lorentz structure, not included here for clarity is associated with the $ZZZZ$ interaction due to the presence in the Feynman gauge of another interaction with four Goldstone bosons.
We refer to Table~5 in ref.~\cite{Chen:2022gxv} for more details.
}
%
%\begin{small}
%\begin{verbatim}
\begin{small}
\begin{lstlisting}[language=Python, backgroundcolor=\color{backcolour}]
particles: [23,23,23,23], # Z Z Z Z
color: [c0 = 1 ],
lorentz: [VVSS1 = Metric(1,2) , VSVS1 = Metric(1,3), 
          VSSV1 = Metric(1,4) , SVVS1 = Metric(2,3), 
          SVSV1 = Metric(2,4) , SSVV1 = Metric(3,4)],
couplings: {(c0, VVSS1): GC_65, (c0, VSVS1): GC_65,
            (c0, VSSV1): GC_65, (c0, SVVS1): GC_65,
            (c0, SVSV1): GC_65, (c0, SSVV1): GC_65},
orders: {QED: 2},
\end{lstlisting}
\end{small}
%\end{verbatim}
%\end{small}

\item[(iii) Multiply propagator factor]

Each of the Lorentz structures of the new interactions will be passed to \ALOHA\ to generate the standard \HELAS\ subroutines. For the subroutines associated with a propagator for a gauge boson, we apply the following transformation to obtain the 5-component FD gauge current from the Feynman gauge current.
Using the gauge vector in eq.\,(\ref{eq:FD}) in the FD gauge, where $q^\mu$ is the four-momentum of the propagating boson with the mass $m$, we convert the output current, \texttt{wf(0:3)} for the gauge boson and \texttt{wf(4)} for the associated Goldstone boson, to the current defined in the FD gauge, \texttt{wfd(0:4)}, as:
\begin{small}
\begin{lstlisting}[
language=Python,
backgroundcolor=\color{backcolour}]
 ci   = (0.d0,1.d0)
 q(4) = -ci * m
 nq   = n(0)*q(0) - n(1)*q(1) - n(2)*q(2) - n(3)*q(3)
 js1  = (n(0)*wf(0)-n(1)*wf(1)-n(2)*wf(2)-n(3)*wf(3)) / nq
 js2  = (q(0)*wf(0)-q(1)*wf(1)-q(2)*wf(2)-q(3)*wf(3) 
&     - dconjg(q(4))*wf(4)) / nq
 wfd(0:4) = wf(0:4)-q(0:4)*js1-n(0:4)*js2
\end{lstlisting}
\end{small}
Here, the four vector, $n^\mu$ in eq.\,\eqref{eq:FD}, is expanded to $n^M$ with $n^4 = 0$.
See Sec.~2.2 as well as Appendix~A in ref.~\cite{Chen:2022gxv} for details.
\end{description}

%%%%%%%%%%%%%%%%%%%%%%%%%%%%%%
%\clearpage
\section{An example in SMEFT}
\label{sec:3}
As a demonstration and a test of the above procedure,
we apply the method in SMEFT with a dimension-6 operator~\cite{Zhang:1994fb,Whisnant:1994fh,Chen:2022yiu,Liu:2023yrb,Barger:2023wbg,Cassidy:2023lwd}
\begin{eqnarray}
 {\cal L} ={\cal L}_{\rm SM}
+ 
\left\{ \frac{C}{\Lambda^2}~Q_3^\dagger t_R \tilde{\phi}
~\left(\tilde{\phi}^\dagger\tilde{\phi}-\frac{v^2}{2}\right) + {\rm h.c.} \right\}
,
\label{eq:SMEFTLag}
\end{eqnarray}
where $Q_3 = (t_L, b_L)^T$ and\footnote{We adopt the
\fr\ notation for the Higgs doublet with hypercharge $1/2$ which differs from the Higgs Lagrangian
given in ref.\,\cite{Chen:2022gxv} by the overall sign of the three
Goldstone bosons.}
\begin{eqnarray}
 \tilde{\phi} =\left ( \frac{v+H-i\pi^0}{\sqrt{2}}, -i\pi^- \right).
\end{eqnarray}
When we take the coefficient as~\cite{Barger:2023wbg}
\begin{eqnarray}
\frac{C}{\Lambda^2}
=
\frac{\sqrt{2}(g_{\rm SM}-ge^{i\xi})}{v^2}
,
\end{eqnarray}
the phenomenological Lagrangian
\begin{eqnarray}
{\cal L}_{ttH} = -g \bar{t} (\cos\xi + i\sin\xi \gamma_5) tH
,
\end{eqnarray}
for CP violating top quark Yukawa coupling is obtained.
Throughout this section, we take
\begin{eqnarray}
 g = g_{\rm SM}^{} = \frac{m_t}{v}
 ,
 \end{eqnarray}
so that the only non-SM parameter is the CP phase, $\xi$.
In the following, $\xi=0$ stands for the SM.

\begin{figure}[t]
\subfigure[]
{\includegraphics[height=0.25\textwidth,clip]{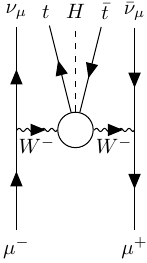}}
\subfigure[]
{\includegraphics[height=0.25\textwidth,clip]{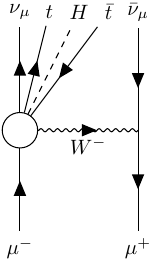}}
\subfigure[]
{\includegraphics[height=0.25\textwidth,clip]{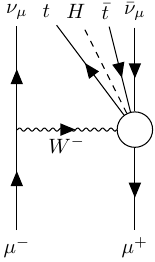}}
\subfigure[]{
\includegraphics[height=0.25\textwidth,clip]{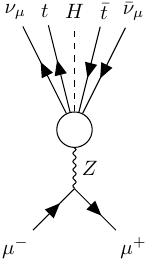}}
\subfigure[]{
\includegraphics[height=0.25\textwidth,clip]{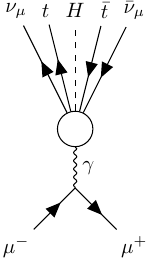}}
\\
\includegraphics[height=0.25\textwidth,clip]{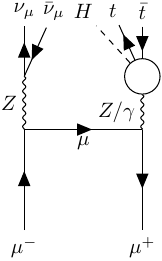}
\includegraphics[height=0.25\textwidth,clip]{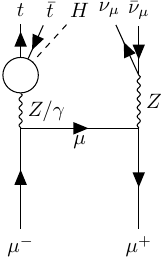}
\includegraphics[height=0.25\textwidth,clip]{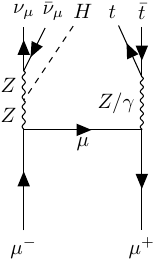}
\includegraphics[height=0.25\textwidth,clip]{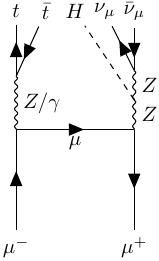}
\\
 \hspace*{0.5cm}(f1)
  \hspace*{2.0cm}(f2)
   \hspace*{2.0cm}(f3)
    \hspace*{2.0cm}(f4)
\caption{The Feynman diagrams for the process 
$\mu^- \mu^+ \to \nu_\mu \bar{\nu}_\mu t \bar{t} H$ are classified into
six groups.  
(a) WWF: $W^-W^+$ fusion;
(b) $\mu^- W^+$: $\mu^- W^+$ fusion; 
(c) $W^-\mu^+$: $W^- \mu^+$ fusion; 
(d) anni-$Z$: $\mu^- \mu^+$ annihilation with $s$-channel $Z$ exchange; 
(e) anni-$\gamma$: $\mu^- \mu^+$ annihilation with $s$-channel $\gamma$ exchange; 
(f1-f4) anni-$\mu$: $\mu^- \mu^+$ annihilation with $t$-channel $\mu$ exchange. }
\label{fig:feynmandiagram}
\end{figure}

By using the method described in section~\ref{sec:2}, we generate
the amplitudes for the process
\begin{eqnarray}
\mu^- \mu^+ \to \nu_\mu \bar{\nu}_\mu t \bar{t} H
,
\label{proc:vmvmttH}
 \end{eqnarray}
 in the tree-level by using \mg.
We find 118 Feynman diagrams in the FD gauge, 
compared to 89 diagrams in the unitary (U) gauge.
The number of diagrams reduce to 89 in the FD gauge and
87 in the U gauge, respectively, in the SM.
In order to study the interference patterns among the
Feynman diagrams, we classify the diagrams into the
following 6 subgroups:
\begin{subequations}
\begin{align}
\begin{split}
&\rm WWF: \rm~ W^-W^+~fusion,
\end{split}
\\
\begin{split}
&\rm \mu^- W^+: \mu^- W^+ ~{\rm~fusion},
\end{split}
\\
\begin{split}
&\rm W^-\mu^+: W^- \mu^+ ~{\rm fusion},
\end{split}
\\
\begin{split}
&{\rm anni}\text{-}Z: \mu^- \mu^+ {\rm annihilation~ with}~ s\text{-}{\rm channel~ }Z{\rm~exchange},
\end{split}
\\
\begin{split}
&        {\rm anni}
\text{-}\gamma: \mu^- \mu^+ {\rm annihilation~ with}~ s\text
{-}{\rm channel}~\gamma {\rm~exchange},
\end{split}
\\
\begin{split}
&{\rm anni}\text{-}\mu:  \mu^- \mu^+ {\rm annihilation~ with}~ t\text{-}{\rm channel}~ \mu~ {\rm exchange},
\end{split}
\end{align}
\label{eq:diagrams}
\end{subequations}
which are illustrated in Fig.\,\ref{fig:feynmandiagram}.
\begin{table}[h]
\begin{tabular}{lccccccccccc}
\hline
\multirow{2}{*}{{\bfseries No.\,of}}&&&\multicolumn{4} {c}{\bfseries SM}&&\multicolumn{4} {c}{\bfseries SMEFT}
\\
 \cline{4-12}
{\bfseries diagrams}&&&{\bfseries U}&&{\bfseries FD}&&&{\bfseries U}&&{\bfseries FD}
\\
\hline
\hline
{a) WWF}&&&19&&21&&&20&&30
\\
\hline
{b) $\mu^- W^+$} &&&11&&11&&&11&&13
\\
\hline
{c) $W^-\mu^+$}&&&11&&11&&&11&&13
\\
\hline
{d) anni-$Z$}&&&24&&24&&&25&&36
\\
\hline
{e) anni-$\gamma$}&&&8&&8&&&8&&10
\\
\hline
{f) anni-$\mu$}&&&14&&14&&&14&&16
\\
\hline
\hline
{\bfseries Total}&&&87&&89&&&89&&118
\\
\hline
\end{tabular}
\caption{ The number of Feynman diagrams given by \mg~for
the process $\mu^- \mu^+ \to \nu_\mu \bar{\nu}_\mu t \bar{t}H$,
with the six types of diagrams.
The four columns give, from left to right,
the SM in the U gauge, the SM in the FD gauge,
the SMEFT model of eq.\,\eqref{eq:SMEFTLag} in the U gauge, and in the FD gauge. 
The numbers represent the number of diagrams
in each category.}
\label{tab: MGcatogery}
\end{table}

In short, the WWF diagrams (a) have two $t$-channel $W$ propagators,
one from the $\mu^- \to \nu_\mu$ leg, the other from the
$\mu^+ \to \bar{\nu}_\mu$ leg. 
The $\mu^- W^+$ diagrams (b) have one $t$-chanel $W$ propagator emitted
from the $\mu^+ \to \bar{\nu}_\mu$ leg, whereas
the $W^-\mu^+$ diagrams (c) have one $t$-chanel $W$ propagator emitted
from the $\mu^- \to \nu_\mu$ leg.
The anni-$Z$ (d) and anni-$\gamma$ (e) diagrams have $s$-channel $Z$ and $\gamma$
exchange, respectively, while the anni-$\mu$ (f) diagrams contain
$t$-channel $\mu$ exchange.

In Table\,\ref{tab: MGcatogery}, we summarize the numbers of Feynman diagrams given by \mg~for each group.
The left column is for the SM, whereas the right column
is for the SMEFT Lagrangian of eq.\,(\ref{eq:SMEFTLag}).
In each column, the left-hand-side gives the diagram numbers in
the U gauge, whereas the right-hand-side is for the
FD gauge. 
%
%%%%%%%%%%%%%%%%%%%%
%Fig2%
\begin{figure}[t]
\subfigure[]{\includegraphics[width=0.45\textwidth,clip]{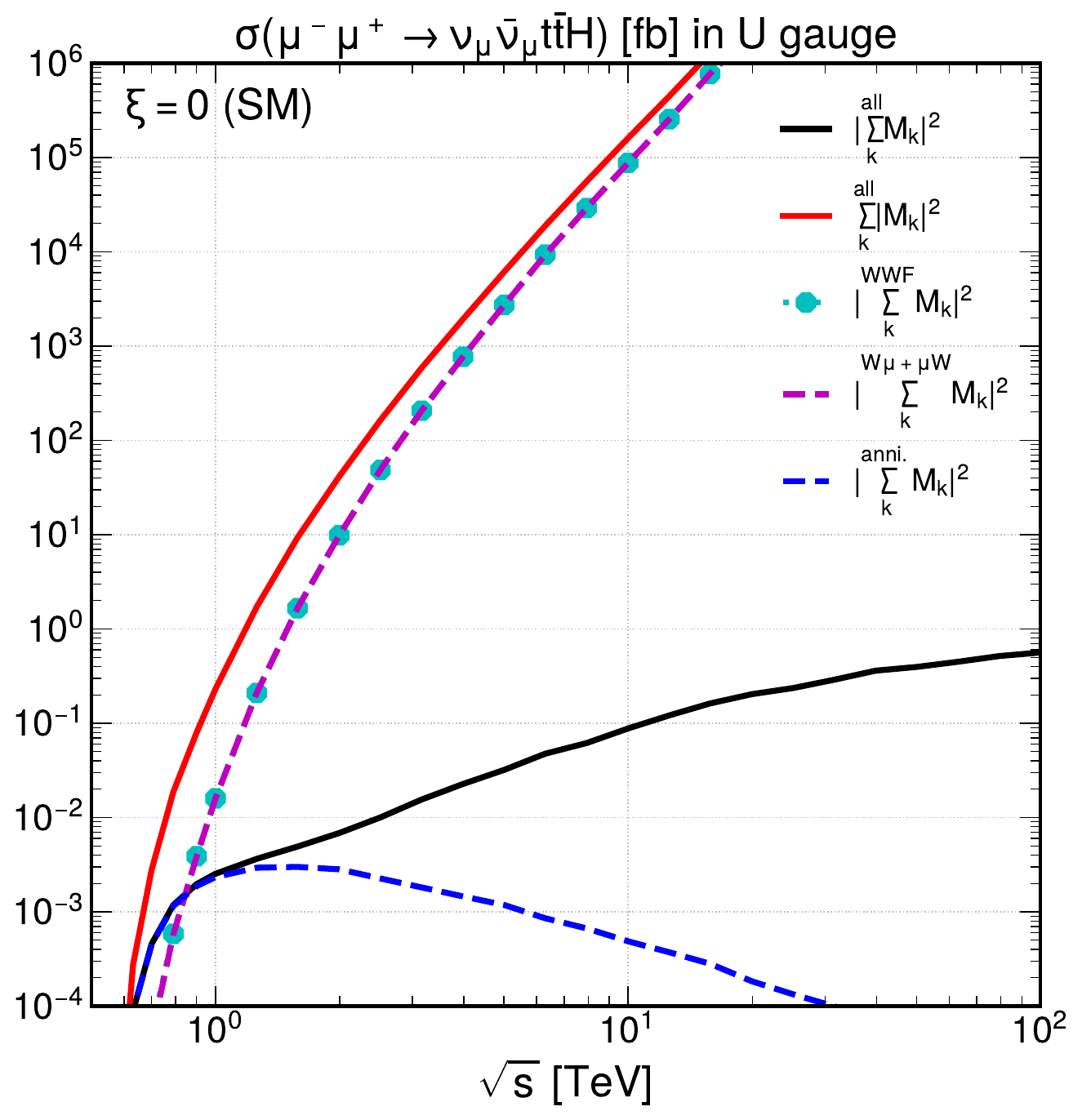}}
\subfigure[]{\includegraphics[width=0.45\textwidth,clip]{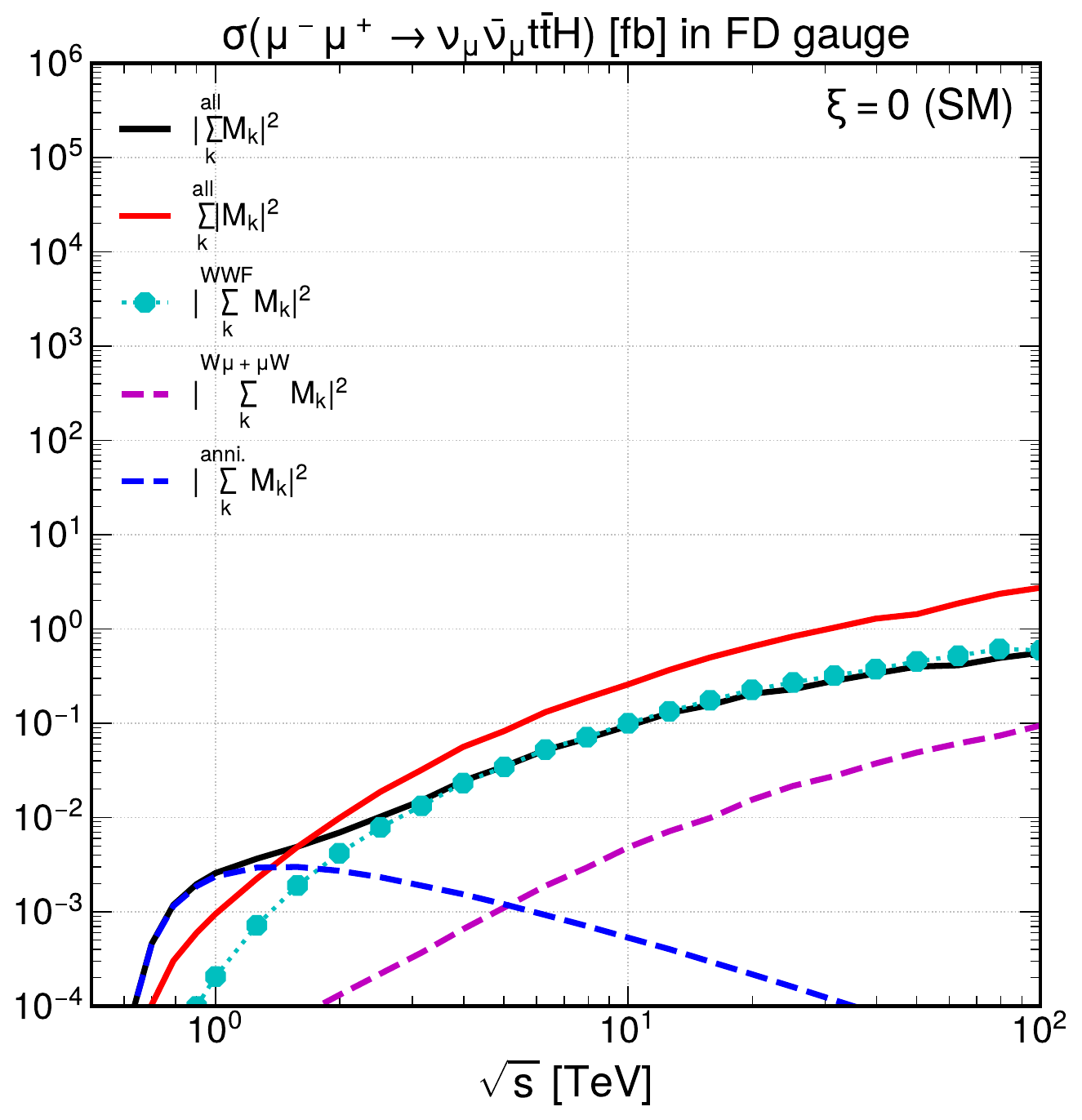}}
\caption{Total cross section of $\mu^-\mu^+\to \nu_\mu \bar{\nu}_\mu t \bar{t} H$ at $\xi=0$ (SM).
The black-solid line denotes the total cross section $|\sum\limits_k^{\rm all}{\cal M}_k|^2$ with 87 diagrams in the  U gauge (a) and with the 89 diagrams in the FD gauge (b). 
The red solid curve denoted as $\sum\limits_k^{\rm all}|{\cal M}_k|^2$ gives the contribution of the total sum of the squares of each diagram.
The cyan dotted line with circle denoted as
$|\sum\limits_k^{\rm WWF}{\cal M}_k|^2$ gives the contribution of $W$ boson fusion diagrams. 
The magenta dashed line denoted as $|\sum\limits_k^{W\mu+\mu W}{\cal M}_k|^2$, gives the contribution of all single $W$ exchange diagrams.
The blue dotted line denoted as $|\sum\limits_k^{\rm anni.}{\cal M}_k|^2$ gives the contribution of all the annihilation type diagrams.
  }
\label{fig:tthvv_SM}
\end{figure}
In Fig.\,\ref{fig:tthvv_SM}, we show the total cross section of the process
eq.(\ref{proc:vmvmttH}) as a function of the total colliding energy $\rts$
in the range $0.5 {\rm~ TeV} < \rts < ~100$ TeV in the SM ($\xi=0$). 
For simplicity, we assume unpolarized muon beams. 
The left panel, Fig.\,\ref{fig:tthvv_SM}(a) is for the U gauge, whereas the right
panel, Fig.\,\ref{fig:tthvv_SM}(b) is for the FD gauge.
The total cross section, given by the black solid curves
are identical between the two panels, showing the gauge
invariance of the total sum of all the Feynman amplitudes, $\sum\limits_k^{\rm all}{\cal M}_k$, and hence their absolute value square, 
\begin{eqnarray}
 | \sum
 \limits_k^{\rm all}{\cal M}_k |^2,
 \label{eq:sumsq}
\end{eqnarray}
which gives the observable cross section
\begin{eqnarray}
    \sigma=\frac{1}{2s}\frac{1}{2}\frac{1}{2}
    \sum\limits_{\rm helicities}
    \int|\sum
 \limits_k^{\rm all}{\cal M}_k|^2
 d\Phi_{\nu_\mu\bar{\nu}_\mu t\bar{t}H},
 \label{eq:cs}
\end{eqnarray}
for unpolarized $\mu$ beams, where $\Phi_{\nu_\mu\bar{\nu}_\mu t\bar{t}H}$ stands for the final states phase space.
Shown by the red solid curves are the total sum of
the absolute square of each diagram,
which is obtained by replacing the term (\ref{eq:sumsq}) by
\begin{eqnarray}
\sum\limits_k^{\rm all}| {\cal M}_k |^2,
\label{eq:sqsum}
\end{eqnarray}
in the cross section formula~(\ref{eq:cs}). Both in eqs.\,(\ref{eq:sumsq}) and (\ref{eq:sqsum}), ${\cal M}_k$ stands for the amplitude of the diagram $k$,
which runs from $k=1$ to 87 in the U gauge, from $k=1$ to 89 in the FD gauge in the SM, as given in the left bottom row of Table\,\ref{tab: MGcatogery}. 
Helicity sum and average,
as well as  the phase space integrals
are done exactly as for the
total cross section.

The ratio of the red solid curve\,\eqref{eq:sqsum} obtained from the sum of the squared amplitudes and
the total cross section
given by the
black solid curve obtained from the square of the total sum of the amplitudes\,\eqref{eq:sumsq} 
\begin{eqnarray}
     R =
     \frac{ \sum\limits_{\rm helicities}\int d\Phi \sum\limits_k |{\cal M}_k|^2 }
{\sum\limits_{\rm helicities} \int d\Phi |\sum\limits_k {\cal }{\cal M}_k|^2 }
,
\label{eq:R}
\end{eqnarray}
is a measure of subtle cancellation among interfering
amplitudes.
As is well known, the red solid curve grows rapidly
with energy in the U gauge, making the ratio $R$ from $92$ at 1 TeV to $1.8\times10^6$ and $6.3\times10^9$ at 10 and 100 TeV, respectively.
In the FD gauge, in contrast, the red solid curve
has the same order of magnitude with the total cross section given by the black solid curve. 
The rate $R$ is found to be $0.37$,~$2.8$~and $4.9$ at $\rts=1,10$ and 100 TeV, respectively.
$R>1$ at $\rts\gtrsim 2$ TeV tells destructive interference among amplitudes.

Also shown in Fig.\,\ref{fig:tthvv_SM} are the partial contribution of
the subsets of diagrams identified in eq.~(\ref{eq:diagrams}) and in Fig.\,\ref{fig:feynmandiagram}, while the number of diagrams in each subset is given in Table\,\ref{tab: MGcatogery}.
%where
The cyan dotted lines with solid circle show the contribution of all the diagrams in the WWF category subgroup (a); 
\begin{eqnarray}
 | \sum\limits_ k^{\rm WWF} {\cal M}_k |^2,
\end{eqnarray}
the magenta dashed curves stand for
the contribution of single $W$ exchange diagrams of the groups (b) and (c);
\begin{eqnarray}
| \sum\limits_k^{W\mu+\mu W} {\cal M}_k |^2,
\end{eqnarray}
and finally the blue dashed curves give the total contribution of all the annihilation diagrams $k$ in the groups (d), (e) and (f);
\begin{eqnarray}
| \sum\limits_k^{\rm anni.} {\cal M}_k |^2.
\end{eqnarray}
We first note that the blue dashed curves are
identical between the U and FD gauges, because the
total sum of (d), (e) and (f) diagrams are identical
to the full amplitudes for the processes,
\begin{eqnarray}
 \mu^- \mu^+ \to \nu_e \bar{\nu}_e t \bar{t} H
,
\label{proc:vevettH}
 \end{eqnarray}
 (or 
 $\mu^- \mu^+ \to \nu_\tau \bar{\nu}_\tau t \bar{t} H$)
 provided that the model satisfies the $\mu-e(\tau)$ universality.
The breakdown of the annihilation amplitudes will be studied at the end of this section.

In the U gauge, Fig.\,\ref{fig:tthvv_SM}(a), both the magenta and cyan curves are far larger than the total
cross section, and that they are degenerate in the
entire energy range.
The two curves are about {$1.0\times10^6$ ($4.4\times10^9$)} times larger than the
total cross section curve at $\rts = 10$~(100) TeV.
These numbers are of the same order of magnitude of the ratio $R$ as defined in eq.\,\eqref{eq:R}
This tells that the subtle cancellation takes place
between the WWF type diagrams (a) and the single
$W$ exchange diagrams (b) and (c) in the U gauge.

In the FD gauge, shown in Fig.\,\ref{fig:tthvv_SM}(b), in contrast, the cyan dotted
curve for the WWF amplitudes saturates the total cross
section at $\rts \gtrsim 3$~TeV.
This agrees with the expectation that the 5-component weak boson representation
gives the weak boson fusion amplitudes with the
physical weak boson PDF, as pointed out first
by Kunszt and Soper in the axial gauge~\cite{Kunszt:1987tk}.
It has been shown in ref.~\cite{Chen:2022xlg} that the FD gauge propagator of the weak
bosons is identical to the axial gauge propagator
of ref.~\cite{Kunszt:1987tk} by taking the LC gauge limit.
The single $W$ exchange diagram contributions,
depicted by the magenta dashed curve, are about
a factor of 5 below the total cross
section at the highest energy of $ \rts\sim 100$~TeV.
The absence of subtle cancellation, 
the saturation of the total cross section by the
WWF type contributions, 
and the dominance of the annihilation contributions
at low energies ($\rts \lesssim 1$ TeV) are all
consistent with a physical picture based on the
weak boson PDF approximation~~\cite{Dawson:1984gx,Kunszt:1987tk,Ruiz:2021tdt}. 
From Fig.\,\ref{fig:tthvv_SM}(b), 
we can tell that the total cross
section is slightly below 
the WWF contribution at highest energies ($\rts\gtrsim 50$ TeV), where the red and black solid curve ratio of $R\sim 3.6$ suggests destructive interference among WWF and single $W$ exchange amplitudes.
Studying further details of the interference patterns among the FD gauge amplitudes is beyond
the scope of the present paper, whose aim is
mainly to demonstrate the validity of the prescription
given in section\,\ref{sec:2}.

\begin{figure}[b]
\subfigure[]{\includegraphics[width=0.45\textwidth,clip]{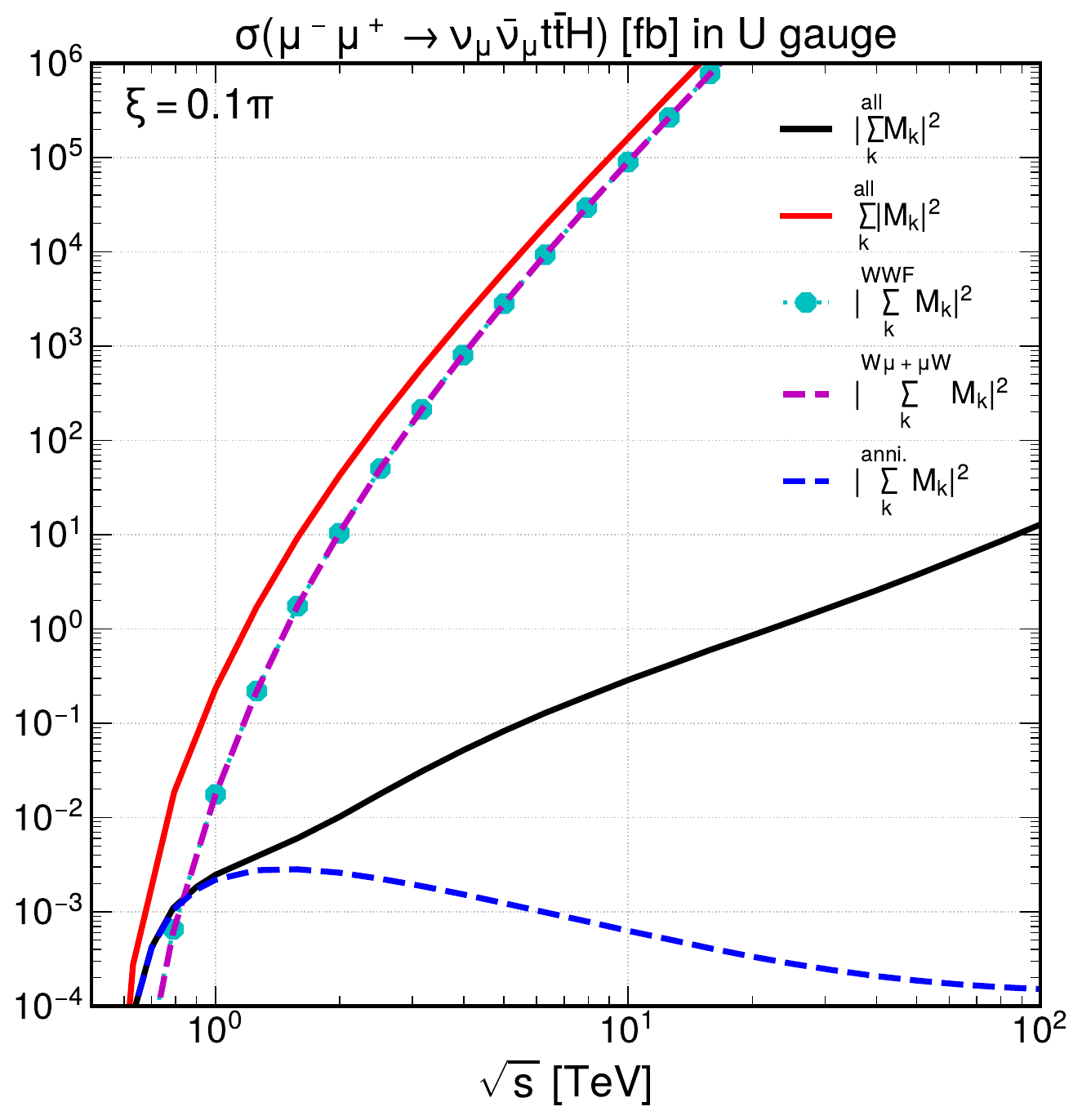}}
\subfigure[]{\includegraphics[width=0.45\textwidth,clip]{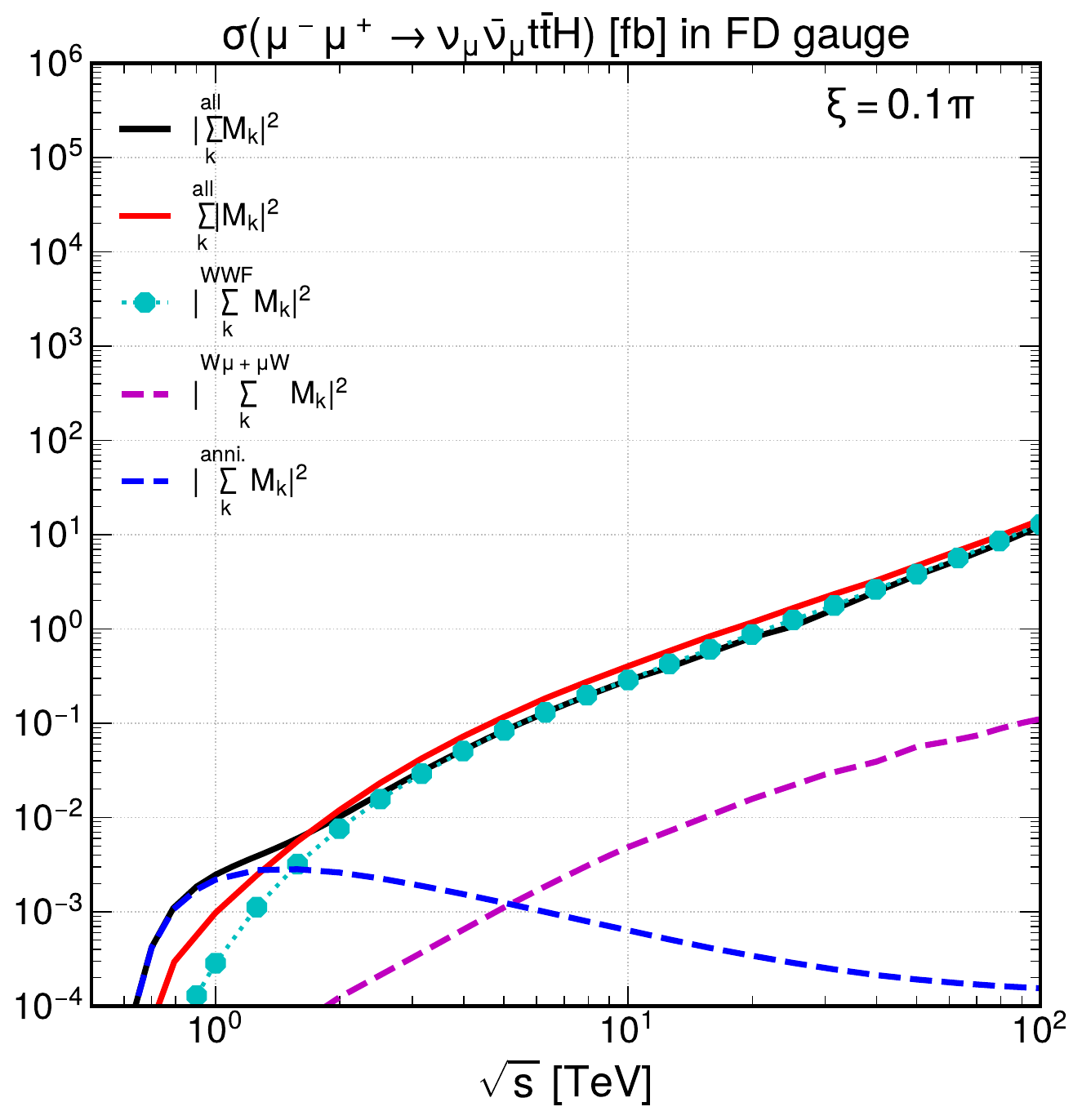}}
\caption{ Cross section of $\mu^-\mu^+\to\nu_\mu\bar{\nu}_\mu t\bar{t}H$ with $\xi=0.1\pi$. The line types are the same as in Fig.\,\ref{fig:tthvv_SM}.}
\label{fig:01pi_U_FD}
\end{figure}

In Figs.\,\ref{fig:01pi_U_FD}(a) and (b), we show the same set of curves
in the presence of the non-SM phase, $\xi = 0.1\pi$.
Both the black solid curve for the total cross section
and the blue dashed curve for the annihilation
contribution are identical between the U gauge (a)
and the FD gauge (b), as expected.
The total cross section (black solid curve) is not
sensitive to the non-SM phase of $|\xi| = 0.1\pi$ at
low energies ($\rts \lesssim 1.2$~TeV), while it
becomes about a factor of 3 times larger than the SM
cross section at $\rts \sim 10$~TeV, about a factor
of 20 times larger at $\rts \sim 100$~TeV.
This has been observed first in ref.~~\cite{Barger:2023wbg} in the U gauge,
and our results in Fig.\,\ref{fig:01pi_U_FD}(b) confirms that we obtain
the same results in the FD gauge, following the
prescription given in section\,\ref{sec:2}.

%%%%%%%%%%%
%Fig3%
Subtle gauge theory cancellation between the contribution
of the $W^-W^+$ fusion diagrams (cyan dotted curve
dabbed WWF) and that of the single weak boson exchange
diagrams (magenta dashed curve dabbed $W^-\mu^++\mu^- W^+$) in the U gauge remains
similar as in the SM case, shown in Fig.\,\ref{fig:tthvv_SM}(a).
This has been expected, since there is only one additional
diagram in the U gauge for the $W^-W^+$ fusion amplitudes,
whose contribution has been evaluated in terms of the $W_L^- W_L^+ \to t\bar{t}H$
amplitudes in ref.~\cite{Barger:2023wbg}. 
Therefore, subtle cancellation in the U gauge takes place between the WWF and single $W$ exchange ($W^-\mu^++\mu^- W^+$) diagrams, just as in the SM.
In fact, we cannot observe significant $\xi$ dependence in the three rapidly growing curves between Fig.\,\ref{fig:tthvv_SM}(a) and Fig.\,\ref{fig:01pi_U_FD}(a). 
The $\xi$ dependence of the total cross section, where the black solid curve in Fig.\,\ref{fig:01pi_U_FD} grows more rapidly with the colliding energy $\rts$ than the SM prediction in Fig.\,\ref{fig:tthvv_SM}, is obtained in the U gauge only after subtle cancellation among amplitudes of huge magnitude.

In a sharp contrast, we can make the following observation by comparing the $\xi=0.1\pi$ results in Fig.\,\ref{fig:01pi_U_FD}(b) and the $\xi=0$ (SM) results in Fig.\,\ref{fig:tthvv_SM}(b) in the FD gauge: the total cross section is dominated by the WWF contribution at high energies, both in the SM and for $\xi=0.1\pi$, and the rising total cross section at high energies for $\xi=0.1\pi$ is due to the rise of the WWF contribution. We do not observe significant $\xi$ dependence in the single $W$ exchange contribution ($W^-\mu^++\mu^-W^+$), shown by the magenta dashed curve both in the Figs.\ref{fig:tthvv_SM}(b) and \ref{fig:01pi_U_FD}(b).

In ref.~\cite{Barger:2023wbg}, the high energy behavior of the total cross
section has been calculated by assuming the dominance of
the longitudinally polarized weak boson fusion subprocess,
\begin{eqnarray}
     W^-_L W^+_L \to t \bar{t} H
     ,
\label{eq:WWttH}
\end{eqnarray}
and then by approximating its cross section by that of 
the corresponding Goldstone collision process,
\begin{eqnarray}
 \pi^- \pi^+ \to t \bar{t} H.
 \label{proc:pipittH}
\end{eqnarray}
The Goldstone boson equivalence theorem~\cite{Cornwall:1974km,Chanowitz:1985hj} tells that
the helicity amplitudes for the above two processes satisfy
\begin{eqnarray}
 {\cal M}\left(W^-_L W^+_L \to t_h \bar{t}_{\bar{h}} H\right)
={\cal  M}(\pi^- \pi^+ \to t_h \bar{t}_{\bar{h}} H)
\left\{ 1 + {\cal O}\left(\frac{m_W^2}{E_W^2}\right) \right\},
\end{eqnarray}
where $h$ and $\bar{h}$ are $t$ and $\bar{t}$ helicities,
respectively. 
In particular, the high energy limit of the helicity amplitudes
for the Goldstone boson collision process (\ref{proc:pipittH}) have been calculated analytically~\cite{Barger:2023wbg} by using the dimension-6 vertex
in the effective Lagrangian (\ref{eq:SMEFTLag}):
\begin{eqnarray}
 {\cal L}_{ttH\pi\pi }
= \frac{g_{\rm SM}-g e^{i\xi}}{v^2}  t_L^\dagger t_R H\pi^+\pi^- + {\rm h.c.}\,.
\label{lag:pipittH}
\end{eqnarray}
The above term gives the only dimension-6 vertex which
contributes to the process (\ref{proc:pipittH}), and the corresponding
amplitudes should dictate the high energy behavior.
This has been confirmed in ref.~\cite{Barger:2023wbg} by comparing the
analytic Goldstone boson amplitudes and the helicity amplitudes of the
WWF subprocess (\ref{eq:WWttH}) evaluated numerically by using \mg\ in the U gauge.
%

%%%%%%%%%
%Fig4%
\begin{figure}[t]
\subfigure[]{\includegraphics[width=0.45\textwidth,clip]{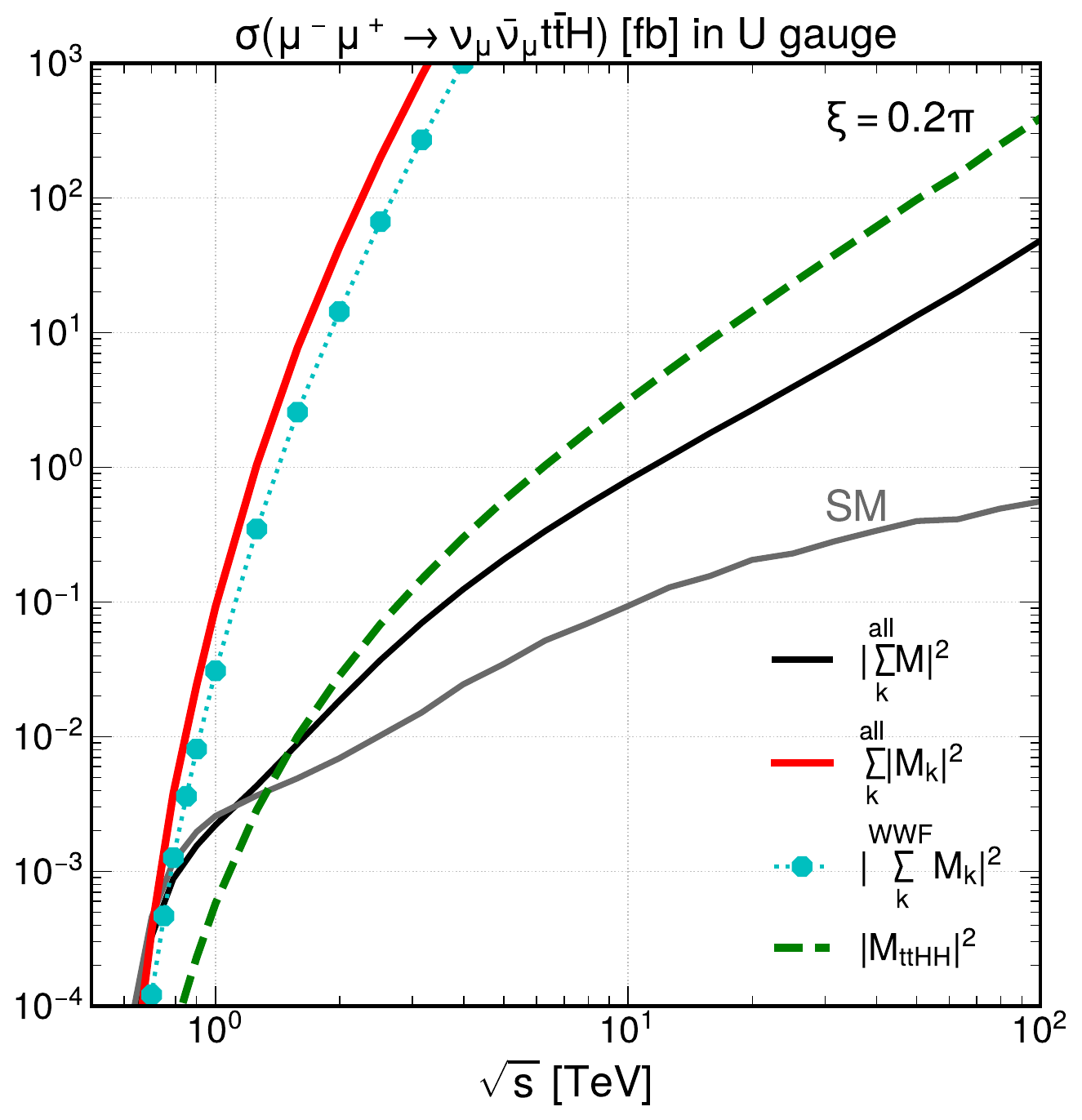}}
\subfigure[]{\includegraphics[width=0.45\textwidth,clip]{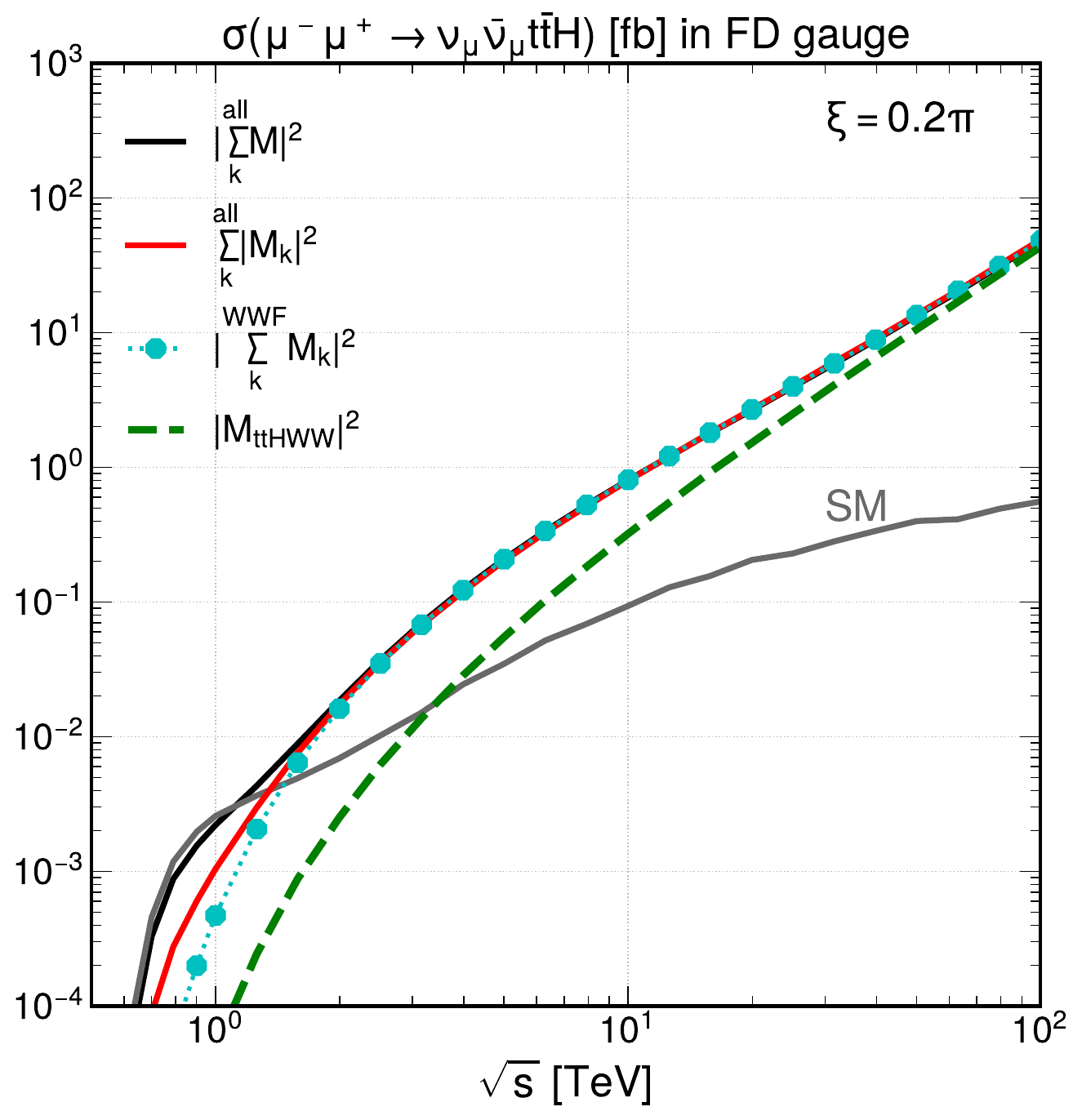}}
\caption{Cross section of $\mu^-\mu^+\to \nu_\mu\bar{\nu}_\mu t\bar{t}H$ with $\xi=0.2\pi$ in the U gauge  (a) and in the FD gauge (b). 
The black solid curve denotes the total cross section.
The red solid curve gives the contribution of the total sum of the squares of each diagram.
The cyan dotted curve gives the contribution of the $W$ boson fusion diagrams.
 The green dashed curve shows the contribution of the single diagram with the contact $ttHH$ vertex, ${\cal M}_{ttHH}$ in the U gauge (a) and that of the diagram ${\cal M}_{ttHWW}$ in the FD gauge (b) with the contact $ttHWW$ vertex. The SM cross section ($\xi=0$) is given by the gray solid line as a reference.
}
\label{fig:vmvm_02pi}
\end{figure}
%%%%%%%%%
%
%%%%%%%
%Fig.5%
\begin{figure}[t]
\subfigure[]
{\includegraphics[width=0.2\textwidth,clip]{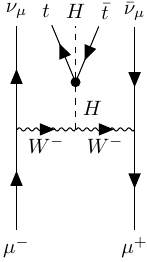}}
\hspace{1.5cm}
\subfigure[]
{\includegraphics[width=0.2\textwidth,clip]{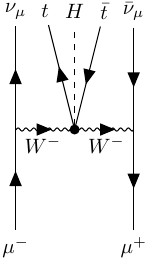}}
\caption{ Feynman diagram with the contact $ttHH$ vertex (a) in both the U and FD gauges and the $ttHWW$ vertex (b) in the FD gauge only.}
\label{fig:feynman47}
\end{figure}
%
%%%%%%%

In Fig.\,\ref{fig:vmvm_02pi}, we show the total cross section for the $\mu^-\mu^+$
collision process (\ref{proc:vmvmttH}) for $\xi=0.2\pi$ by the black solid curve. 
As a reference, the SM prediction ($\xi=0$) is given by the grey
solid curve.
In addition to the red solid curve for the sum of squares of
each amplitude and the cyan dotted curve for the WWF contribution, in Fig.\,\ref{fig:vmvm_02pi}(b)
we show by green dashed curve the contribution of the dimension-6 vertex (\ref{lag:pipittH}) in the FD gauge, $|{\cal M}_{ttHWW}|^2$, depicted as the Feynman diagram Fig.\,\ref{fig:feynman47}(b).
It is clearly seen from Fig.\,\ref{fig:vmvm_02pi}(b) that the total cross section
is dominated by the WWF contribution at $\rts\gtrsim 3$~TeV,
and then it is dominated by $|{\cal M}_{ttHWW}|^2$ at $\rts\gtrsim 100$~TeV,
in the FD gauge.
This is an example of the property of the FD gauge amplitudes,
where the Goldstone boson equivalence is manifest~\cite{Wulzer:2013mza,Chen:2016wkt}.

Shown in Fig.\,\ref{fig:vmvm_02pi}(a) are the results for the U gauge.
As in Figs.\,\ref{fig:tthvv_SM}(a) and \,\ref{fig:01pi_U_FD}(a), both the sum of squares of each
amplitude (red solid) and the WWF contribution (cyan dotted)
grow rapidly with energy, while the physical cross section
(black solid) is exactly the same as the FD gauge.
In the figure, we show by green dashed curve the contribution
of the single Feynman diagram of Fig.\,\ref{fig:feynman47}(a), which has the
$ttHH$ vertex,
\begin{eqnarray}
 {\cal L}_{ttHH}
= \frac{3(g_{\rm SM}-g e^{i\xi})}{2v} t_L^\dagger t_R H^2 + {\rm h.c.}\,,
\end{eqnarray}
whose mass dimension is 5. 
In the U gauge, the above $ttHH$ vertex is the only interactions
whose mass dimension is larger than 4.
The green dashed curve grows with energy faster than the
total cross section shown by the black solid curve, and
is about a factor of 8 larger than the total cross section
at $\rts \sim 100$~TeV.
At this energy, the red solid curve gives $3.9\times10^{9}$~fb,
or about $8.0\times10^{7}$ times larger than the total cross section.
This again confirms the findings of ref.~\cite{Barger:2023wbg}, where the
amplitudes of the diagram Fig.\,\ref{fig:feynman47}(a), or its WWF subamplitudes
have been evaluated analytically.

Summing up, the FD gauge amplitudes are free from subtle
gauge cancellation among interfering diagrams, and the
total cross section for the process (\ref{proc:vmvmttH}) is dominated by
the weak boson fusion (WWF) subamplitudes at high energies,
both in the SM, and with nonzero CP phase $\xi$.
The Goldstone boson equivalence between the longitudinally
polarized weak boson and its associated Goldstone boson
is manifestly realized in the FD gauge amplitudes. 
In addition, we find that the high energy behavior of
the FD gauge amplitudes are dictated by the amplitudes
with the highest dimensional vertex.

In the rest of this section, we examine the annihilation
amplitudes, whose contribution given by blue dashed curves
in Figs.\,\ref{fig:tthvv_SM} and \ref{fig:01pi_U_FD} are identical between the U gauge and
the FD gauge.
This simply reflects the fact that the `annihilation'
diagrams consist of a gauge invariant subset of the
amplitudes, which corresponds to the full set of Feynman
diagrams for a certain physical process such as the process in eq.\,(14),
where the final neutrino flavor in the process (\ref{proc:vmvmttH}) is 
changed from $\nu_\mu$ to $\nu_e$ (or $\nu_\tau$).
All the non-annihilation diagrams are then forbidden
by the muon number conservation in the SM.

\begin{figure}[t]
\subfigure[]
{\includegraphics[width=0.45\textwidth,clip]{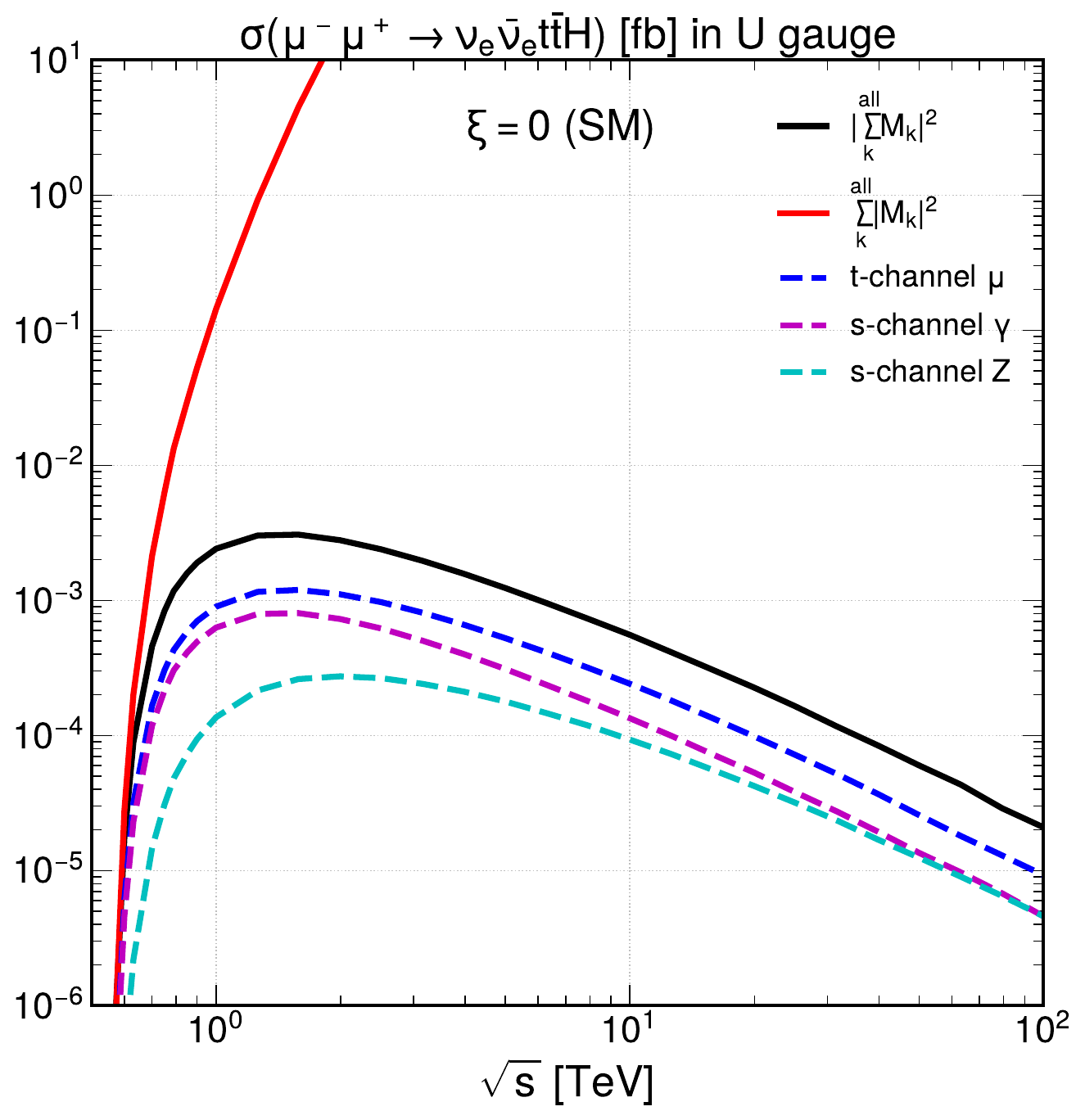}}
\subfigure[]
{\includegraphics[width=0.45\textwidth,clip]{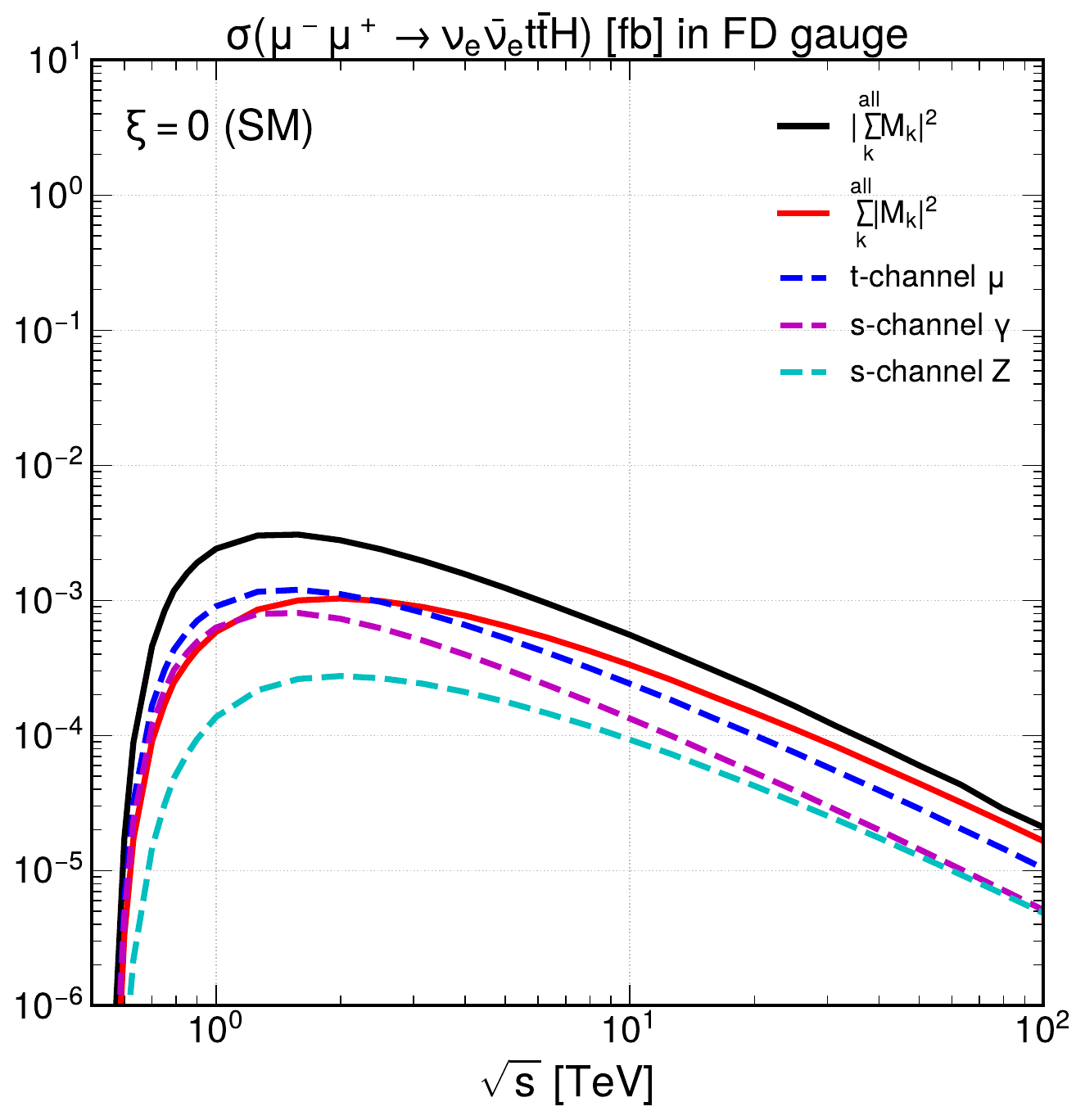}}
\caption{Cross section of $\mu^-\mu^+\to\nu_e\bar{\nu}_e t\bar{t}H$ for $\xi=0$ (SM) for (a) U gauge, and (b) for FD gauge. The black solid line is the total cross section. The red solid line is for the $\sum\limits_k^{\rm all}{|\cal M}_k|^2$, summing up the squared of each amplitude. The blue dashed line is summing up the $t$-channel $\mu$ exchange diagram. The magenta dashed line is summing up the photon exchange diagram. The cyan dashed line is summing up the $s$-channel $Z$ exchange diagrams.}
\label{fig:vevettHSM}
\end{figure}

In Fig.\,\ref{fig:vevettHSM}, we show the total cross section of the
process $\mu^-\mu^+\to \nu_e\bar{\nu}_e t\bar{t}H$, eq.\,(\ref{proc:vevettH}) as a function of the colliding muon pair
energy, in the U gauge (a) and in the FD gauge (b).
In addition to the total cross section given by the
black solid line, we show the contributions of the 
three subgroups, those of the $t$-channel muon exchange
diagrams in blue dashed curves, those of the $s$-channel
$\gamma$ exchange diagrams in magenta dashed curves,
and the $s$-channel $Z$ exchange diagrams by cyan
dashed curves.
We find that not only the total cross section given
by the black solid curves, but also all the three
subgroups of the amplitudes give exactly the same
cross section in Figs.\,\ref{fig:vevettHSM}(a) and (b).
Although there is no physical process in the SM which is given
by the diagrams of the three annihilation subgroups,
we can retain the $s$-channel $\gamma$ or $Z$ exchange
diagrams only by changing the $SU(2)\times U(1)$
quantum numbers of the muon.
We can hence regard the subgroups of Feynman diagrams
as the full set of diagrams for the annhilation of
such exotic leptons.
%%%%%%%%%%%%%%%%%%%%%%
%%%%Fig.6%%%%%%%
%%%%%%%%%%%%%%%%%%%%%%

The total sum of the squares of each Feynman
amplitude is shown by red solid curves.
We observe rapid growth of the red curve in the U gauge,
as shown in Fig.\,\ref{fig:tthvv_SM}(b) for the process (\ref{proc:vmvmttH}).
The ratio $R$ of the red and black solid curves in Fig.\,\ref{fig:vevettHSM}(a),
is about 60 at $\rts=1$ TeV, which grows rapidly to about $1.0\times10^{8}$ at $\rts=10$~TeV,
and $5.9\times10^{13}$ at $\rts=100$~TeV, which 
grows even faster with energy than
what we find for the process\,(\ref{proc:vmvmttH}) in Fig.\,\ref{fig:tthvv_SM}(a).

In Fig.\,\ref{fig:vevettHSM}(b), we find that the red solid curve is
consistently below the black solid curve for the
total cross section, where the ratio $R$ of
eq.\,(\ref{eq:R}) is about 
$0.24$, $0.60$ and $0.79$, 
at $\rts=1,~10$, and $100$~TeV,
respectively.
The $R$ value below unity tells an overall constructive 
interference among Feynman amplitudes.
We find that the total cross section (black solid curve)
is approximately the sum of the three subamplitude 
contributions.

%%%%%%%%%%%%%%%%%%%%
%%Fig7
\begin{figure}[b]
\subfigure[]
{\includegraphics[width=0.45\textwidth,clip]{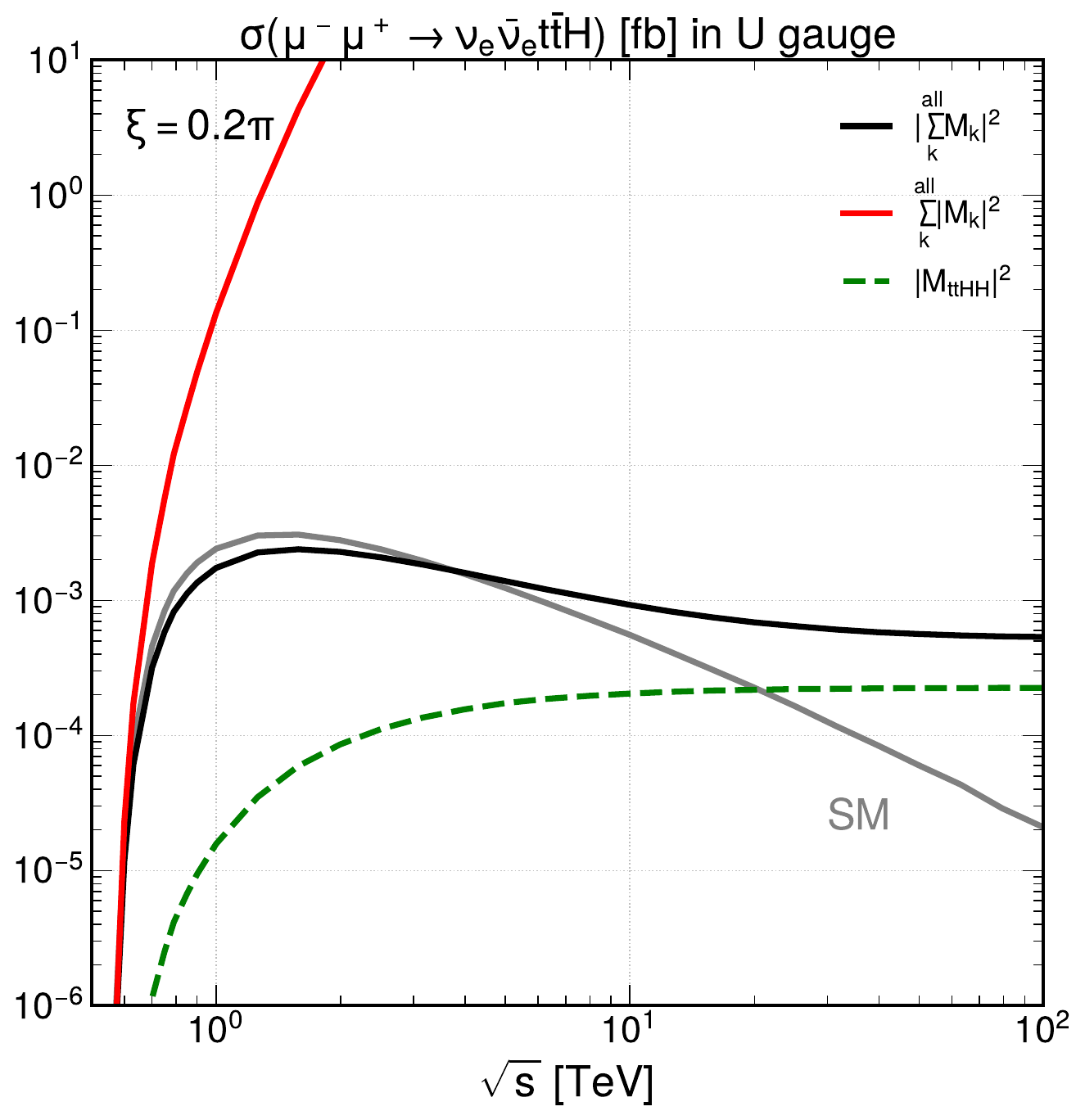}}
\subfigure[]
{\includegraphics[width=0.45\textwidth,clip]{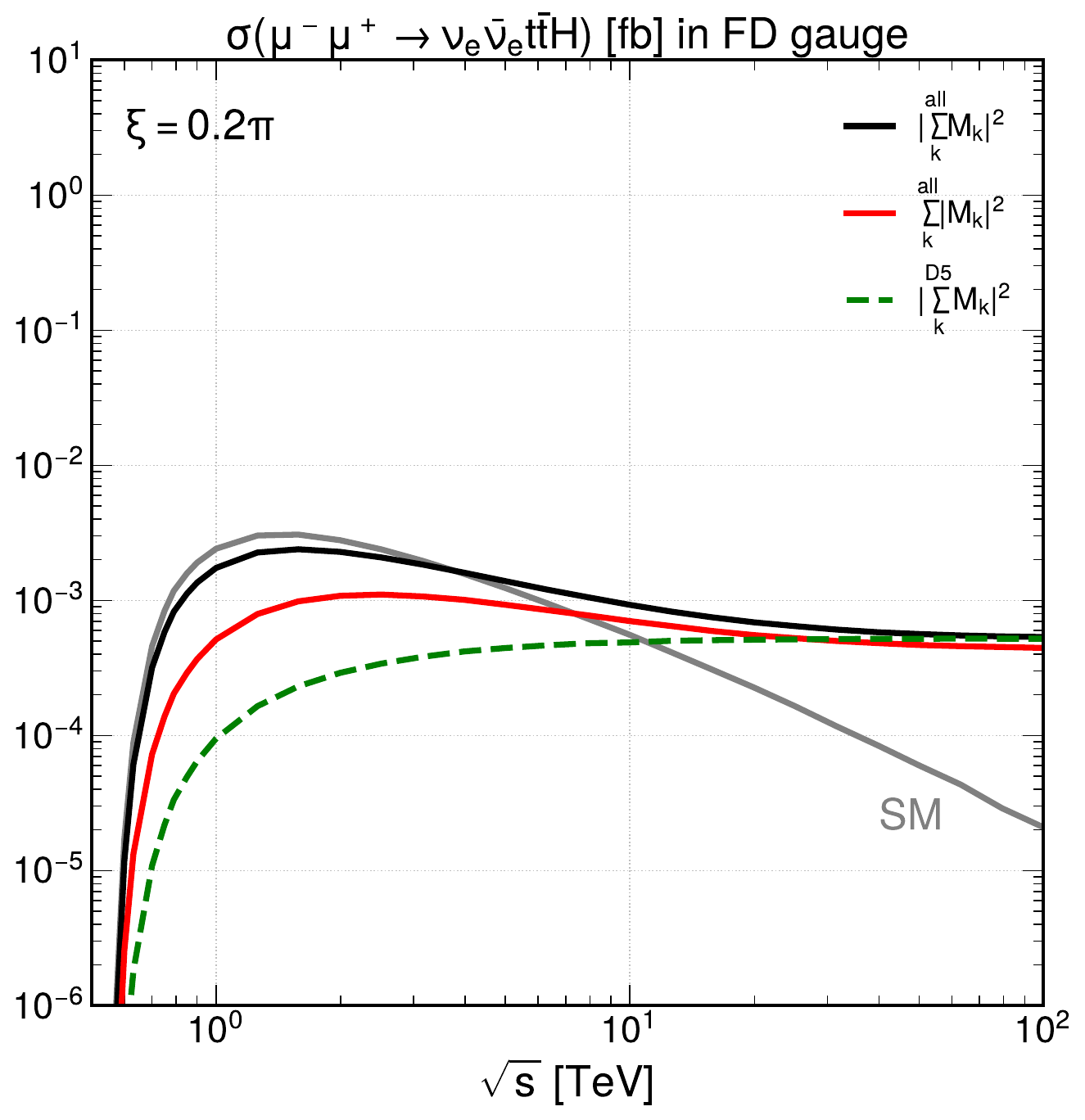}}
\caption{Cross section of $\mu^-\mu^+\to\nu_e\bar{\nu}_e t\bar{t}H$ for the $\xi=0.2\pi$ in the U gauge (a) , and in the FD gauge(b). The line types are the same as Fig.\,\ref{fig:vmvm_02pi}, where the green dashed curve shows the contribution of the diagram with the dimension-5 $ttHH$ vertex in the U gauge (a), while it shows the contribution of all the diagrams with dimension-5 vertices $ttHH$, $ttHZ$ and $ttZZ$ in the FD gauge (b). } 
\label{fig:vevettH02pi}
\end{figure}
%%%%%%%%%%%%%%%%%%%%
Finally, in Figs.\,\ref{fig:vevettH02pi}(a) and \ref{fig:vevettH02pi}(b), we show the results
for our effective Lagrangian model (\ref{eq:SMEFTLag}), when the
CP phase value is $\xi=0.2\pi$.
The total cross section is given by the black
solid curves, while we also show the SM ($\xi=0$) results
from Fig.\,\ref{fig:vevettHSM} by grey solid curves, which are both
exactly the same between in the U gauge (a) and
the FD gauge (b). 
The ratio $R$ of the red and black solid curves,
eq.\,(\ref{eq:R}), grows rapidly with energy in the U gauge,
from $77$ at 1~TeV, to $6.1\times10^7$ at 10 TeV and $2.3\times10^{12}$ at 100~TeV,
respectively.
These values are similar to the SM case reported
above for Fig.\,\ref{fig:vevettHSM}(a), and hence the interference
pattern in the U gauge is not affected significantly
by the presence of the non-SM interactions in our
effective Lagrangian model of eq.\,(\ref{eq:SMEFTLag}).

The $\xi$ dependence of the total cross section has been reported systematically in ref.~\cite{Barger:2023wbg}. It can also
be learned from comparing the black and grey solid
curves, which are common between Figs.\,\ref{fig:vevettH02pi}(a) and (b).
The total cross section for $\xi=0.2\pi$ is slightly
smaller than the SM cross section at $\xi=0$ below
$\rts\sim 3$~TeV, which confirms the trend observed in refs.\,\cite{Gunion:1996vv,BhupalDev:2007ftb,Ananthanarayan:2014eea,Hagiwara:2017ban,Ma:2018ott,Azevedo:2022jnd,Cheung:2023qnj} for the process
\begin{eqnarray}
 e^- e^+ \to t \bar{t} H,
\label{proc:eettH}
 \end{eqnarray}
since the process (\ref{proc:vevettH}) can be regarded as a $Z$ boson
emission correction to the process (\ref{proc:eettH}), and hence
the $\xi$ dependence should be similar at low
energies.
At high energies, $\rts \gtrsim 5$~TeV, the black
curve stays above the grey curve for the SM cross
section, and decreases very slowly with energy,
reaching about $5\times 10^{-4}$~fb at 100~TeV.
Because the total cross section decreases with
energy in the SM, the cross section is about 1.7 and 26 larger than the SM
at $\rts=10$ and $100$~TeV, respectively. 
The asymptotically constant behavior of the
total cross section suggests contribution of
higher dimensional interactions in the
effective Lagrangian model (\ref{eq:SMEFTLag}).
%
%%%%%%%%%%%%%%%%%
%%%%%Fig.8%%%
\begin{figure}[b]
\subfigure[]
{\includegraphics[height=0.35\textwidth,clip]{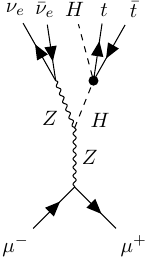}}
\hspace{1.0cm}
\subfigure[]
{\includegraphics[height=0.35\textwidth,clip]{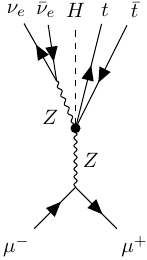}}
\caption{ Feynman diagram with the  contact $ttHH$ vertex (a) in both gauges and the that with $ttHZZ$ vertex in the FD gauge (b).}
\label{fig:ttHZZ}
\end{figure}
%%%%%%%%%%%%%%%%%
%
Shown in Fig.\,\ref{fig:ttHZZ}(a) is the Feynman diagram with
the $ttHH$ vertex, whose mass dimension is 5.
In the U gauge, where all the Goldstone boson
vertices are inactive, this is the only diagram
with higher dimensional vertex.
We show by green dashed curve the contribution of
the diagam Fig.~\ref{fig:ttHZZ}(a) in the U gauge, labeled
as $|{\cal M}_{ttHH}|^2$.
The single diagram contribution given by the green
curve grows with energy and has similar energy
dependence with the black curve at highest energies,
reaching about $40\%$ of the total cross section at $\rts=100$ TeV.
The red solid curve for the total sum of squares
of each amplitude grows rapidly with energy just
as in Fig.\,\ref{fig:vevettHSM}(a) for the SM.
Therefore, subtle gauge cancellation in the U
gauge is not affected significantly by the
presence of the non-SM interactions in the
effective Lagrangian (\ref{eq:SMEFTLag}).

From Table\,\ref{tab: MGcatogery}, we have 47 Feynman diagrams for
the process (\ref{proc:vevettH}) in the U gauge.
One of these 47 diagrams is the one given by
Fig.\,\ref{fig:ttHZZ}(a), ${\cal M}_{ttHH}$, and the remaining 46 diagrams don't
contain higher dimensional vertices.
The results shown in Fig.\,\ref{fig:vevettH02pi}(a) for the U gauge tell us,
e.g.\ at $\rts=100$~TeV,
the total sum of the 46 Feynman diagrams
gives, after subtle gauge cancellation,
the amplitudes whose magnitude is similar to $|{\cal M}_{ttHH}|$.
Although the energy and the kinematical dependences of $|{\cal M}_{ttHH}|^2$ can be understood from the analytic expression of the diagram Fig.\,\ref{fig:ttHZZ}(a), that the
total sum of the remaining 46 amplitudes give similar amplitudes at high energies tells
us a miraculous aspect of gauge cancellation in the U gauge.

In Fig.\,\ref{fig:ttHZZ}(b), we show the Feynman diagram with
the $ttHZZ$ vertex, which is the only dimension-6
vertex in the effective Lagrangian of eq.\,(\ref{eq:SMEFTLag}). 
Because the corresponding diagram with the dimension-6
$ttHWW$ vertex in Fig.\,\ref{fig:feynman47}(b) has been found to
dominate the total cross section of the process
(\ref{proc:vmvmttH}) at high energies, as shown by green dashed curve
in Fig.\,\ref{fig:vmvm_02pi}(b), we first study the contribution
of the diagram Fig.\,\ref{fig:ttHZZ}(b), whose amplitude can
be denoted as ${\cal M}_{ttHZZ}$. 
We find that
\begin{eqnarray}
    {\cal M}_{ttHZZ} = 0
    ,
\end{eqnarray}
for all helicities at all energies.
We attribute the cause of the vanishing amplitudes
as a consequence of the fact that the virtual $Z$
boson produced from massless lepton pair is purely
transverse, and have no longitudinally polarized
component. 
As a support of this observation, we find that all
the amplitudes with higher dimensional operators
whose Goldstone boson leg is connected directly
to the initial state lepton pair are zero.
In Table\,\ref{tab: MGcatogery}, we have in total 62 diagrams contributing
to the process (\ref{proc:vevettH}), among which 16 diagrams have
higher dimensional vertices, 1 at dimension-6, that
of Fig.\,\ref{fig:ttHZZ}(b), and the remaining 15 diagrams with
dimension-5 vertices.
There are 3 types of dimension-5 vertices,
$ttHH$ in Fig.\,\ref{fig:ttHZZ}(a),
$ttHZ$, and $ttZZ$.
Feynman diagrams with $ttHZ$ and $ttZZ$ vertices
where the vertex is connected by the FD gauge
propagator directly to the initial $\mu^-\mu^+$
current give zero amplitudes.
The remaining 10 diagrams with dimension-5 vertices
give non-zero amplitudes, 1 with $ttHH$,
8 with $ttHZ$, and 1 with $ttZZ$ vertices.
Since all these amplitudes with one dimension-5
vertex are expected to obey the same energy
scaling law, we study the total sum of all
the non-vanishing amplitudes with the
dimension-5 vertices:
\begin{eqnarray}
|\sum\limits_{k}^{D5} {\cal M}_k|^2
,
\end{eqnarray}
whose contribution is shown by the green dashed
curve in Fig.\,\ref{fig:vevettH02pi}(b).
As expected, the green dashed curve merges with
the black solid curve at high energies,
$\rts \gtrsim 50$~TeV.
The total amplitudes in the FD gauge are dominated
by the amplitudes with higher dimensional vertices
as in the case of the amplitudes for the process
(\ref{proc:vmvmttH}) shown in Fig.\,\ref{fig:vmvm_02pi}(b).
The red curve remains slightly below the black
and green curves,  giving the ratio
$R \sim0.83$ at $\rts \sim 100$~TeV in Fig.\ref{fig:vevettH02pi}(b).
The constructive interference is observed among
the 10 non-vanishing diagrams with dimension-5 vertices.
This is in contrast to the case of the total
cross section of the process (\ref{proc:vmvmttH}) shown in Fig.\,\ref{fig:vmvm_02pi}(b), 
where the ratio $R\sim 1$ at high energies,
reflecting the single diagram dominance.

%%%%%%%%%%%%%%%%%%%%%%%%%%%%%%%
\section{Summary}
\label{sec:4}
In this paper, we show how tree-level scattering amplitudes
in the Feynman-Diagram (FD) gauge~\cite{Chen:2022gxv, Chen:2022xlg} can be generated
automatically in gauge models with non-Standard Model (SM)
interactions. 
\begin{itemize}
  \item 
  We start from the toolbox \fr~\cite{Alloul:2013bka}, which can generate 
Feynman rules in the 't Hooft-Feynman gauge automatically
for an arbitrary gauge model with spontaneous symmetry
breaking.
\item 
Based on the universal \fr\ output (\UFO)~\cite{Degrande:2011ua,Darme:2023jdn} in the Feynman gauge,
we introduce 5-component representation of the weak bosons
and their $5\times 5$ propagators automatically in \mg~\cite{Alwall:2014hca}. 
\item 
External 5-component weak boson polarization vectors and
the $5\times 5$ weak boson propagators are common for
all the gauge models, and we adopt the representations
given in the original FD gauge papers~\cite{Chen:2022gxv, Chen:2022xlg}.
\item 
Feynman diagrams for an arbitrary tree-level scattering
amplitude with the 5-component weak bosons and their
propagators can be generated on the flight by \mg.
\item 
All the vertex functions with and among 5-component weak
bosons are obtained by assembling those of the 4-component
weak bosons and the Goldstone bosons, which are generated
automatically by using \ALOHA~\cite{deAquino:2011ub} for an arbitrary gauge
model.
\end{itemize}
The numerical codes for each Feynman diagram generated 
by the above procedure give helicity amplitudes in
the FD gauge.

The above procedure has been explained in detail in
section~\ref{sec:2}. 
In section~\ref{sec:3}, we demonstrate its validity by generating 
the helicity amplitudes for the muon collider process
$\mu^- \mu^+ \to \nu_{\mu} \bar{\nu}_\mu t \bar{t} H$
in an SMEFT~\cite{Leung:1984ni,Buchmuller:1985jz,Grzadkowski:2010es} model with just one dimension-6
operator giving complex top Yukawa coupling~\cite{Barger:2023wbg}.
118 Feynman diagrams are generated in the FD gauge,
as compared to the 89 diagrams in the unitary gauge.
After summing over all the diagrams, we find exact
agreement between the FD and the unitary gauges.

As in the case of the SM processes reported in
refs.~\cite{Hagiwara:2020tbx,Chen:2022gxv, Chen:2022xlg}, the FD gauge amplitudes are free
from subtle gauge cancellation among Feynman
diagrams at high energies,
thus providing an
efficient basis for numerical evaluation of the
amplitudes.
A new finding in this paper is that in the FD gauge,
the high energy behavior of individual Feynman amplitude
is dictated by the mass dimension of the contributing
new physics vertex.
In the weak boson fusion (WWF) amplitudes, the single
diagram with a dimension-6 $ttHWW$ vertex dominates the total
cross section at extreme high energies $\rts \sim 100$~TeV.
Among the $\mu^- \mu^+$ annhilation amplitudes, the
corresponding amplitude with a dimension-6 $ttHZZ$ vertex
is found to vanish, and the 10 non-vanishing amplitudes
with a dimension-5 vertex ($ttHH$, $ttHZ$, $ttZZ$) jointly
dictate the high energy amplitudes at $\rts\gtrsim 50$~TeV.

The above example tells that the FD gauge amplitudes
are useful in identifying the subamplitudes with non-SM
interactions, because individual Feynman amplitude satisfies
the naive scaling law; $n$-point amplitudes scale as $E^{n-4}$
when all the vertices have mass dimension 4.
The amplitude scales like $E^{n-3}$ with one
dimension-5 vertex, or $E^{n-2}$ with one dimension-6 vertex, in the example studied in section~\ref{sec:3}.
This is an additional merit of the FD gauge, because 
the scaling law of the tree-level amplitudes is not
manifest in covariant gauges, including the unitary gauge.

\section*{Acknowledgement}
We are grateful to Vernon Barger for enlightening discussions. This work is supported in part by JSPS KAKENHI Grant No.\,21H01077 and 23K03403. KH is supported by the US Japan Cooperation Program in High Energy Physics.

\bibliography{F2FD}
\bibliographystyle{utphys}

\end{document}